\documentclass[fleqn,usenatbib]{mnras}

\usepackage[T1]{fontenc}

\DeclareRobustCommand{\VAN}[3]{#2}
\let\VANthebibliography\thebibliography
\def\thebibliography{\DeclareRobustCommand{\VAN}[3]{##3}\VANthebibliography}

\usepackage{graphicx}	
\usepackage{amsmath}	
\usepackage{amssymb}	
\usepackage{float}
\usepackage{caption}
\usepackage{subcaption}
\usepackage{gensymb}
\usepackage{placeins}
\usepackage{newtxtext,newtxmath}

\title[Photo-z's using Capsule Networks]{Photometric Redshifts from SDSS Images with an Interpretable Deep Capsule Network}
\author[Dey et al.]{Biprateep Dey$^{1}$\thanks{E-mail:biprateep@pitt.edu},
Brett H. Andrews$^{1}$,
Jeffrey A. Newman$^{1}$,
Yao-Yuan Mao$^{2}$\thanks{NASA Einstein Fellow}, 
\newauthor Markus Michael Rau$^{3,4}$ and 
Rongpu Zhou$^{5}$ \\
\\
$^{1}$Department of Physics and Astronomy and PITT PACC, University of Pittsburgh, Pittsburgh, PA 15260, USA \\
$^{2}$Department of Physics and Astronomy, Rutgers, The State University of New Jersey, Piscataway, NJ 08854, USA \\
$^{3}$Department of Physics, McWilliams Center for Cosmology, Carnegie Mellon University, Pittsburgh, PA 15213, USA\\
$^{4}$High Energy Physics Division, Argonne National Laboratory, Lemont, IL 60439\\
$^{5}$Lawrence Berkeley National Laboratory, 1 Cyclotron Road, Berkeley, CA 94720, USA
}

\date{Accepted XXX. Received YYY; in original form ZZZ}

\pubyear{2021}

\begin{document}
\label{firstpage}
\pagerange{\pageref{firstpage}--\pageref{lastpage}}
\maketitle

\begin{abstract}
Studies of cosmology, galaxy evolution, and astronomical transients with current and next-generation wide-field imaging surveys like the Rubin Observatory Legacy Survey of Space and Time (LSST) are all critically dependent on estimates of photometric redshifts. Capsule networks are a new type of neural network architecture that is better suited for identifying morphological features of the input images than traditional convolutional neural networks. We use a deep capsule network trained on $ugriz$ images, spectroscopic redshifts, and Galaxy Zoo spiral/elliptical classifications of $\sim$400,000 Sloan Digital Sky Survey (SDSS) galaxies to do photometric redshift estimation.  We achieve a photometric redshift prediction accuracy and a fraction of catastrophic outliers that are comparable to or better than current methods for SDSS main galaxy sample-like data sets ($r\leq17.8$ and $z_{\mathrm{spec}}\leq0.4$) while requiring less data and fewer trainable parameters. Furthermore, the decision-making of our capsule network is much more easily interpretable as capsules act as a low-dimensional encoding of the image. When the capsules are projected on a 2-dimensional manifold, they form a single redshift sequence with the fraction of spirals in a region exhibiting a gradient roughly perpendicular to the redshift sequence. We perturb encodings of real galaxy images in this low-dimensional space to create synthetic galaxy images that demonstrate the image properties (e.g., size, orientation, and surface brightness) encoded by each dimension. We also measure correlations between galaxy properties (e.g., magnitudes, colours, and stellar mass) and each capsule dimension. We publicly release our code, estimated redshifts, and additional catalogues at \url{https://biprateep.github.io/encapZulate-1}.
\end{abstract}
\begin{keywords}
galaxies: distances and redshifts -- methods: statistical -- methods: data analysis
\end{keywords}

\section{Introduction}\label{sec:intro}

Wide-field extra-galactic sky surveys collect photometric or spectroscopic measurements to create 3-dimensional maps of the Universe by measuring on-sky positions and redshifts of a variety of astronomical objects. These maps help us study the growth of the Universe and its large-scale structure over time by measuring various observable quantities as a function of redshift. For example, \citet{Hubble1929HubbleLaw} studied distances to nearby galaxies as a function of redshift to discover the expansion of the Universe and more recently, \citet{ReissEtal1998DarkEnergy} and \cite{PerlmutterEtal1999DarkEnergy} studied the relationship between luminosity distances of Type Ia supernovae and their redshifts to discover cosmic acceleration and hence dark energy. Detection of Baryon Acoustic Oscillations (BAO) using large redshifts surveys \citep{Cole2005BAO, Eisentstein2005BAO} similarly gave us another independent measurement of the cosmic acceleration and other parameters of the concordance model of cosmology. Cosmological redshifts are a proxy for the distance to extra-galactic objects thereby allowing us to measure their intrinsic properties (like luminosity, mass, star formation rate, etc.) and enabling studies of the formation and evolution of galaxies. Accurate redshift measurements of satellite galaxies in the nearby Universe allow us to study the nature and distribution of dark matter and help us constrain models of galaxy formation and evolution. Redshift measurements also help with rapid identification of host galaxies of transient sources for follow-up as made evident by the recent discovery of gravitational wave sources with electromagnetic counterparts \citep{AbbottEtal2017GravWaveEMCounter}. 

Given the long exposure times required and the limited multiplexing of spectroscopic instruments, high precision spectroscopic redshifts (spec-$z$'s) can only be measured for a tiny fraction of galaxies for which we have images. For example, it will be possible to measure spectroscopic redshifts for less than 1\% of the galaxies that will be used in the Rubin Observatory Legacy Survey of Space and Time (LSST) studies of galaxy evolution and cosmology \citep{LSSTEtal2009LSSTDescription}. Because of this limitation, it will be necessary to infer redshift information using imaging data alone; the resulting measurements are called photometric redshifts or photo-$z$'s. Accurate photo-$z$ estimates along with well-calibrated uncertainties will be crucial to achieve the ambitious science goals set for the next generation of photometric surveys like LSST.

Most photo-$z$ estimation methods involve finding a non-linear mapping between photometrically observed properties of galaxies (like apparent magnitudes and colours) and redshift. This is achieved either by fitting the observed photometry with redshifted templates of galaxy spectral energy distributions (SED) (e.g., LePhare, \citealt{Arnouts1999PhotozTemplate}, \citealt{Ilbert2006LephareTemplateFitting}, \citealt{Ilbert2009LephareTemplateFitting}; BPZ, \citealt{Benitez2000PhotozTemplate}; ZEBRA, \citealt{Feldman2006ZebraTemplateFitting}; EAZY, \citealt{Brammer2008PhotozTemplate}; Phosphoros, \citealt{Apostolakos2019PhosphorosTemplateFitting}, Paltani et al. in prep.; MAGPHYS, \citealt{Battisti2019MagphysPhotoz}; \citealt{Lee2020TemplateFitting}) or using a machine learning (ML) based model trained on galaxies with spectroscopic redshifts to approximate this relationship. The optimal method generally depends on the amount and quality of data available and the scientific questions to be addressed. Template-based methods work well for deep, high redshift surveys where the faintness of the galaxies and the broad redshift range covered makes it prohibitively expensive to collect large data sets. However, SED templates often rely on assumptions on galaxy physics (like star formation history or initial mass function), have incomplete coverage of the entire wavelength range and model dust attenuation poorly, all of which are significant sources of errors \citep{SalvatoEtal2019PhotozReview}. 

In contrast, in regimes with more complete training data like that provided by shallow low redshift spectroscopic surveys, common statistical methods like linear regression (e.g., \citealt{Connolly1995ColorRedshift,Beck2016Photoz}) or classical machine learning techniques such as decision trees and random forests (e.g., \citealt{Carliles2010RFPhotoz,Dalmasso2020FlexcodePhotoz,Zhou2021DESIPhotoz,Li2022Photoz}), support vector machines (e.g., \citealt{Wadadekar2005SVMPhotoz,Jones2017SVM}), $K$-nearest neighbours (e.g., \citealt{Ball2008KNNz,Graham2018CMNN}), self-organized mapping (e.g., \citealt{Carrasco2014SOMz,Geach2012SOMphotoz,Wright2020SOM,Myles2021SOMPzDESphotoz}), Gaussian processes (e.g., \citealt{Way2009GPPhotoz,Almosallam2016GPz}) and simple neural networks (e.g., \citealt{Firth2003ANNPhotoz,Tagliaferri2003NNPhotoz, Collister2004ANNz,Cavuoti2017METAPHORPhotoz,Razim2021ANNPhotoz}) tend to outperform template-based methods \citep{Hildebrandt2010PHATChallenge, Schmidt2020LSSTPhotoz,Desprez2020EuclidChallenge}.

A challenge for photo-$z$ estimation methods that take magnitudes and colours as inputs is that there is not enough information available to break various degeneracies in the colour--redshift relation. One way to break these degeneracies is to include information about morphology, orientation, surface brightness, ratios of magnitudes, or visual appearance in general (e.g., \citealt{Stabenau2008SurfaceBrightnessPhoto-z,Jones2017SVM,Gomes2018GPz,Disanto2018PhotozFeatureselection,Nakoneczny2021PhotozFeature}). A galaxy may appear red not just because its stellar population is intrinsically red but because it is a dusty edge-on spiral galaxy. Moreover, the fact that farther objects appear to be smaller and fainter to an observer also give us an additional piece of information to help break degeneracies. Most traditional methods for quantifying galaxy morphology, like ellipticity and S\'ersic index, cannot fully encode all of the visual information that an image of a galaxy provides and hence methods that use images of galaxies directly as inputs (e.g. \citealt{PasquestEtal2019Photoz,Hayat2021PhotozSelfSupervised,Schuldt2021HSCPhotoz,Henghes2022CNNPhotoz}) and rely on artificial neural networks are the current state-of-the-art.

Artificial neural networks are mathematical models, originally developed to mimic the logical operations of the human brain. The simplest unit of such a model (also called an artificial neuron) is a linear transformation of an input followed by some non-linear function (also called an activation function). Successive layers of such transformations arranged together form a deep neural network. The process of training such a model involves finding a set of parameters (also called weights) for these transformations which will minimise a loss function. The optimisation is generally done using the back propagation algorithm \citep{LeCun1985Backprop, Rumelhart1986BackProp} or some optimiser based on it like the Adam optimiser \citep{Kingma2015Adam}. The simplest deep neural network architecture called multi-layer perceptrons or fully connected (FC) networks use successive matrix and non-linear transformations to connect every input feature to an output. A sufficiently deep or wide fully connected network can be used to approximate any function \citep{Cybenko1989UniversalApproximationTheorem, Hornik1989UnivAppTheo,Hornik1991UnivAppTheo} and hence can be used to effectively predict photometric redshifts.

If the input data are images, then the number of trainable weights required for a fully connected neural network architecture becomes very large, making them very inefficient to train and prone to over-fitting to the training data. Convolutional Neural Networks (CNNs; \citealt{FUKUSHIMA1982CNN, LeCun1989CNN}), on the other hand, perform convolution operations using filters whose parameters are learned. Since the same set of filters are reused by stepping across the input images, it reduces the number of trainable parameters. Moreover, each successive convolution layer can extract more complex features which in turn increases the model accuracy while reducing the complexity of the model. Various multi-layered neural network architectures (i.e., deep neural networks; \citealt{LeCun2015DLReview}) built using CNNs have been used to make state-of-the-art photo-$z$ prediction algorithms as they can leverage the pixel level data to extract additional information thereby achieving even better prediction accuracy. \citet{Hoyle2016Photoz} modified the ImageNet challenge-winning AlexNet \citep{Krizhevsky2012AlexNet} to $griz$ images of $\sim$64,000 SDSS galaxies, finding comparable accuracy to the best tree-based classical machine learning algorithms. \citet{Disanto2018Photoz} combined a CNN and a mixture density network to produce photo-$z$ Probability Density Functions (PDFs) generated using Gaussian mixture models and achieved comparable performance to existing efforts in the literature. As larger training data sets become available along with advances in Graphical Processing Unit (GPU) hardware and associated software, training CNNs have become very efficient and currently form the backbone of most state-of-the-art photo-$z$ algorithms. \citet{PasquestEtal2019Photoz} produced the current best photo-$z$’s using a supervised algorithm for the SDSS Main Galaxy Sample, which consists of $\sim$ 500,000 $ugriz$ images with spec-$z$’s in the range $z=0-0.4$. They applied an innovative deep CNN that included five inception modules \citep{Szegedy2015Inception, Szegedy2015Inception2} which use multiple filter sizes within the CNN operating at the same level rather than being stacked sequentially to capture information on different scales efficiently. Recently, self-supervised learning-based approaches have shown promising results on astronomical data sets (e.g., \citealt{Stein2021SelfSupervised}, \citealt{Sarmiento2021SelfSupervised}). \citet{Hayat2021PhotozSelfSupervised} used a self-supervised training scheme paired with a ResNet50-based CNN \citep{HeEtal2016Resnet50} to achieve similar results but with less data. They pre-trained their network on a very large unlabelled data set to find similarities between different augmentations of the inputs and then fine-tuned the network to predict photometric redshifts. When fine-tuned using the whole SDSS main galaxy sample, they achieve state-of-the-art results.

Deep neural network-based methods are continuing to improve but have substantial limitations in terms of the interpretability of the features learnt from images and efficiency in training. Models with larger number of trainable parameters not only require more data and computational resources to train but also are prone to over-fitting. To alleviate some of these issues, we explore the use of a modern deep learning method called capsule networks \citep{Hinton2011CapsNet} to jointly predict photo-$z$'s and basic morphological types of galaxies (spiral/elliptical). Capsule networks are robust to rotations and invariant to viewpoint---a useful quality for analysing randomly oriented galaxies and require less training data and trainable parameters than CNNs because they generalise much better to novel viewpoints \citep{Mazzia2021EfficientCapsNet}. Capsule networks also learn a low-dimensional representation of the input images, which provides us with a way to interpret the features learnt by the model.

In this work we will focus on predicting photo-$z$ point estimates but ideally we would like to quantify the uncertainty in our estimates by predicting full photo-$z$ probability density functions (PDFs). However, producing properly calibrated photo-$z$ PDFs remains extremely challenging. PDFs predicted by neural networks are often poorly calibrated (see e.g., \citealt{Guo2017NNCalibration}) and provide very misleading uncertainty estimates. Moreover, most methods currently used to check the quality of photo-$z$ PDFs (like distributions of probability integral transform, etc.) focus on checking the calibration of the entire sample of PDFs (i.e. global calibration) rather than focusing on the calibration of individual PDFs (i.e. local/individual calibrations). \citet{Amaro2019PhotozComparison} and \citet{Schmidt2020LSSTPhotoz} show that such metrics can  be optimised by pathological but non-physical photo-$z$ PDFs. \citet{Zhao2021CDEDiagnostics} show that global calibration of PDFs does not imply local calibration and proposes new diagnostics which may be used to check for local calibration.  In a future paper, we plan to extend our methods and produce locally calibrated PDFs following the procedure described in \citet{Dey2021Recalibration,Dey2022Recal}. That being said, the prediction errors on our photo-$z$ point estimates are sufficiently small that we can safely use these estimates for studies of the evolution of galaxies, their connection with dark matter halos, and the localisation of transient sources. where photo-$z$ PDFs are not strictly required. 

The paper is organised as follows. In Sec.~\ref{sec:data}, we discuss the various data sets used in this work. In Sec.~\ref{sec:capsnets} we introduce the concept of capsule networks and explain our network architecture. In Sec.~\ref{sec:training} we describe the process of training a multi-task capsule network. In Sec.~\ref{sec:results} we present our results for photo-$z$ point estimates and compare our results with other similar works. We also provide interpretations of the features learnt by the capsule network in order to predict photo-$z$'s. Lastly, in Sec.~\ref{sec:summary} we summarise our results.

\section{Data} \label{sec:data}

\subsection{SDSS Imaging and Spectroscopic Redshifts}
To train and test our models, we use the same pre-processed images and spectroscopic redshifts that were used by \citet{PasquestEtal2019Photoz} for their CNN-based photo-$z$ estimation method and were generously made publicly available by the authors\footnote{\href{https://deepdip.iap.fr/\#item/60ef1e05be2b8ebb048d951d}{https://deepdip.iap.fr/\#item/60ef1e05be2b8ebb048d951d}}. The data set contains 516,525 galaxies with de-reddened $r$ band petrosian magnitudes, $r\leq 17.8$, and spectroscopic redshifts, $z\leq0.4$ selected from the 12th Data Release (DR12) of the Sloan Digital Sky Survey (SDSS; \citealt{Gunn1998SDSSCamera,Gunn2006SDSSTelescope,YorkEtal2000Sdss,Smee2013SDSSSpectrograph,AlamEtal2015SdssDR12}). The sample is mainly defined by the magnitude limit as the redshift limit removes only a few tens of galaxies. The 12th data release of SDSS was used as that was the most recent data release available when this work began. Moreover, there has not been any changes to the data for this particular set of galaxies since DR8, making all SDSS data releases since DR8 equivalent for our purposes.

For this set of galaxies, \citet{PasquestEtal2019Photoz} pre-processed the raw 5 band images after obtaining them from the 8th Data Release of SDSS \citep{Aihara2011SDSSDR8}. They stacked and re-sampled the images to a common $64\times64\times5$ pixel grid centred on the spectroscopic target. The images were background subtracted and photometrically calibrated with the same zero point \citep{Blanton2011SDSSBackground,Padmanabhan2008SDSSCalibration}. No foreground/background objects were removed. Most of the galaxies had only 1 or 2 imaging frames per band, whereas galaxies in Stripe 82 \citep{Jiang2014Stripe82} had up to 64 imaging frames. So, the Stripe 82 galaxies which satisfy our magnitude and redshift cuts defining the parent sample have significantly less noise than the other images. The Stripe 82 galaxies form less than 4\% of the entire data set and can be used to check how amount of noise in the images affect our methods (see Sec.~\ref{sec:point-estimates}). All the galaxies in the data set are spatially resolved, so their sizes, surface brightnesses, morphologies in each band, and the presence of neighbouring and background galaxies provide additional information not captured in spatially integrated photometry. A detailed description of the image processing steps can be found in Sec.~2.1 of \citet{PasquestEtal2019Photoz}. The processed images along with their spectroscopic redshifts used in this work are publicly available.

\subsection{Galaxy Zoo-1 Morphological Class Labels} \label{sec:data-morpho}
We use a deep capsule network to jointly predict the basic morphology of a galaxy along with its redshift. We use crowd-sourced morphological class labels of galaxies from the Galaxy Zoo-1 project \citep{LintottEtal2011GalaxyZoo1} to train our capsule network. Galaxy Zoo-1 labels galaxies as spirals (with various sub-classes), ellipticals, mergers, or stars-and-artefacts. The classifications are considered ``confident'' only if the de-biased fraction of votes received for a class is greater than 0.8. Since the numbers of mergers and artefacts in the images of the SDSS-MGS are very low, we use the spiral and elliptical classes only. This gives us high-quality morphological classifications for 177,442 of the galaxies in our parent data set. We generate morphological class labels for the remaining 339,083 galaxies in our data set, using an iterative semi-supervised system where we train a deep capsule network using the confident class labels and use it to generate the labels for all other galaxies (see Sec.~\ref{sec:morphology-classification} for details). Out of these 339,083 galaxies that do not have a confident classification, we obtain the fraction of votes received in Galaxy Zoo-1 for each class for 296,767 galaxies which we use to cross-check our results. For the remaining 42,316 galaxies, no morphology information was available since they did not pass some of the quality cuts imposed by Galaxy Zoo-1. We do not use these galaxies to asses the quality of our morphological class prediction and only their deep capsule network generated class labels are used for redshift prediction.

\subsection{Catalogue of Galaxy Properties} \label{sec:data-galaxy-prop}
To interpret the features learnt by our deep capsule network, we measure correlations between the low-dimensional encodings of the input images produced by the capsules and various other galaxy properties (see Sec.~\ref{sec:dist-corr}). 
For this purpose, we created a cross-matched catalogue of various observed and estimated physical properties for the galaxies in our data set.

For all galaxies, we query their model magnitudes, composite model (cmodel) magnitudes\footnote{\href{https://www.sdss.org/dr12/algorithms/magnitudes/}{https://www.sdss.org/dr12/algorithms/magnitudes/}} and extinction due to Milky Way dust from \citet{Schlegel1998DustMap} for each of the five SDSS photometric bands from the SDSS DR12 database \citep{AlamEtal2015SdssDR12}. We also query the velocity dispersion ($\sigma_{v}$) measured from the spectra. We use the extinction corrected cmodel magnitudes as a measure of the galaxy magnitudes and use extinction corrected model magnitudes to calculate the colours of the galaxies. We also query measurements of stellar mass ($\mathrm{M}_{\star}$), star formation rate ($\mathrm{SFR}$), and specific star formation rate ($\mathrm{sSFR}$) from the Max Planck Institute for Astrophysics and the Johns Hopkins University (MPA-JHU) value-added catalogue\footnote{\href{https://www.sdss.org/dr12/spectro/galaxy_mpajhu/}{https://www.sdss.org/dr12/spectro/galaxy\_mpajhu/}} available as a part of SDSS DR12. These measurements are based on the methods developed in \citet{Kauffmann2003MPAJHU}, \citet{Brinchmann2004MPAJHU}, and \citet{Tremonti2004MPAJHU}. For estimates of absolute magnitudes ($\mathrm{M}_{u/g/r/i/z}$) we use the measurements from the New York University Value Added Galaxy aCatalog\footnote{\href{http://sdss.physics.nyu.edu/vagc/}{http://sdss.physics.nyu.edu/vagc/}} (NYU-VAGC; \citealt{Blanton2005NYUVAGC}) for objects common between our data set and the NYU-VAGC within a tolerance of 1 arcsecond. We also use measurements of S\'ersic-index in the $r$ band ($n_{r}$) and the corresponding 90\% light radius ($R_{90,r}$) from the NYU-VAGC as a proxy for a galaxy's size.
 
A small number of objects in our data set do not have matches with the external catalogues and there are also some measurements in these catalogues that are problematic. We only use the objects in our data set that have cross matches with the external catalogues for each of the galaxy properties. We also remove measurements of any property which are more than 5 units of median absolute deviation (scaled to replicate Gaussian standard deviation) away from the median of that property. This step is done to remove the small number ($<1\%$) of problematic measurements of galaxy properties that can affect our analysis.

\section{Capsule Networks}\label{sec:capsnets}
Convolutional Neural Networks (CNNs; \citealt{FUKUSHIMA1982CNN, LeCun1989CNN}) are currently the de facto standard for neural network architectures when the input data are images. They work by learning weights for a set of convolutional filters which extract useful features from the images. As the filters are reused by translating them across the input, CNNs have fewer trainable parameters compared to their fully connected counterparts and also invariant to small translations of the object of interest with respect to the background. Each successive layer of the deep network extracts more and more complex features in an hierarchical fashion. CNNs have been immensely successful in solving problems in computer vision (e.g., \citealt{Krizhevsky2012AlexNet,Szegedy2015Inception,liu2022ConvNext} ) and have been used extensively for predicting photometric redshifts from images (e.g., \citealt{Hoyle2016Photoz,Disanto2018Photoz,PasquestEtal2019Photoz,Hayat2021PhotozSelfSupervised,Henghes2022CNNPhotoz}).

Though CNNs are invariant to translations by design \citep{ Lecun1998CNN, Lee2009CNN}, they use pooling layers (i.e. replacing the input with the local maximum or average value) to locally combine the signal and reduce dimensionality \citep{Ranzato2007CNN}. This comes at the cost of losing precise location and pose information (see e.g., \citealt{Hinton2011CapsNet, Hinton2021PartWhole}). To solve this problem, \citet{Hinton2011CapsNet} proposed that artificial neural networks should be organised as local groups that perform complex computations on their inputs and encapsulates the results into highly informative output vectors. These vector counterparts of artificial neurons are called capsules and the entire computational chain is termed as a capsule network. Each capsule vector should learn to recognise the presence of a visual entity irrespective of its orientation, viewing conditions, etc. They should not only encode the probability of the object being present but also encode a set of ``instantiation parameters'' for the entity (e.g., location, size, orientation, colour, etc.). For an ideal capsule network, the encoded probability of an object being present should stay the same but the instantiation parameters should change when the input image goes through some transformation (like, rotation, translation, occlusion, etc.).

Though \citet{Hinton2011CapsNet} introduced the idea of a capsule network, a concrete architecture and training methodology was not proposed. More recently, \citet{SabourEtal2017Capsnet} proposed a training method called the dynamic routing algorithm which made capsule networks viable. Their architecture encodes the ``probability'' of an object being present using the length of the capsule vectors. During the training process, information from each capsule is weighted before passing it onto the next layer of capsules via the dynamic routing algorithm \citep{SabourEtal2017Capsnet}. The elements of the transformation matrices between two successive capsules are determined by the gradient descent algorithm whereas the routing weights are determined so as to maximise the cosine similarity (i.e., vector dot product) between the capsule vectors of the two consecutive layers in an iterative fashion. Dynamic routing allows capsule networks to focus on specific sections or traits of the input data while making decisions. After each routing step, the capsules are scaled using the nonlinear squashing function, $f(\mathbf{v}) = \frac{\lVert \mathbf{v} \rVert^2}{1 + \lVert \mathbf{v} \rVert^2}\frac{\mathbf{v}}{\lVert \mathbf{v} \rVert}$ which re-scales the length of each capsule to be between 0 and 1 and acts as the nonlinear activation function for the layer. 

The original implementation of capsule networks in \citet{SabourEtal2017Capsnet} was geared towards the classification of grey-scale handwritten digits. The same implementation was adapted for an astronomical application by \citet{Katebi2019CapsNetMorpho} for morphological classification of galaxies, both of which are easier problems compared to photo-$z$ estimation. Consequently, they got state-of-the art results while using only a single layer of capsules and a routing algorithm that does not train efficiently if multiple capsule layers are present. To do well in more complicated tasks, it is helpful to have multiple layers of capsules (i.e., a deep capsule network). For this work we adopt the deep capsule network architecture and dynamic routing algorithm as proposed in \citet{RajasegaranEtal2019Deepcaps}. They propose convolution operation based capsule network layers and a 3D-convolution based routing algorithm which reduces the number of trainable parameters and makes the routing process significantly more efficient thereby making deep capsule networks possible. They also use skip connections \citep{HeEtal2016Resnet50} which add outputs of earlier layers with the outputs of layers ahead of it to improve the convergence of the training process by preventing the gradients from vanishing and allowing information from earlier capsules to flow efficiently to later ones. \citet{RajasegaranEtal2019Deepcaps} also introduced an improved class independent decoder network which reconstructs the input image from the final layer capsules and thereby enforces that the components of the capsule vectors form a low-dimensional encoding of the input image. The class-independent nature of the decoder ensures that the capsule dimensions encode the same properties for both morphological classes. A mathematical description of the capsule network layers and routing algorithms mentioned in this section is given in Appendix~\ref{appendix:routing}.

\subsection{Our Capsule Network Architecture}
\begin{figure*}
  \includegraphics[trim={2.5cm 0 0 0},clip, width=1.14\textwidth]{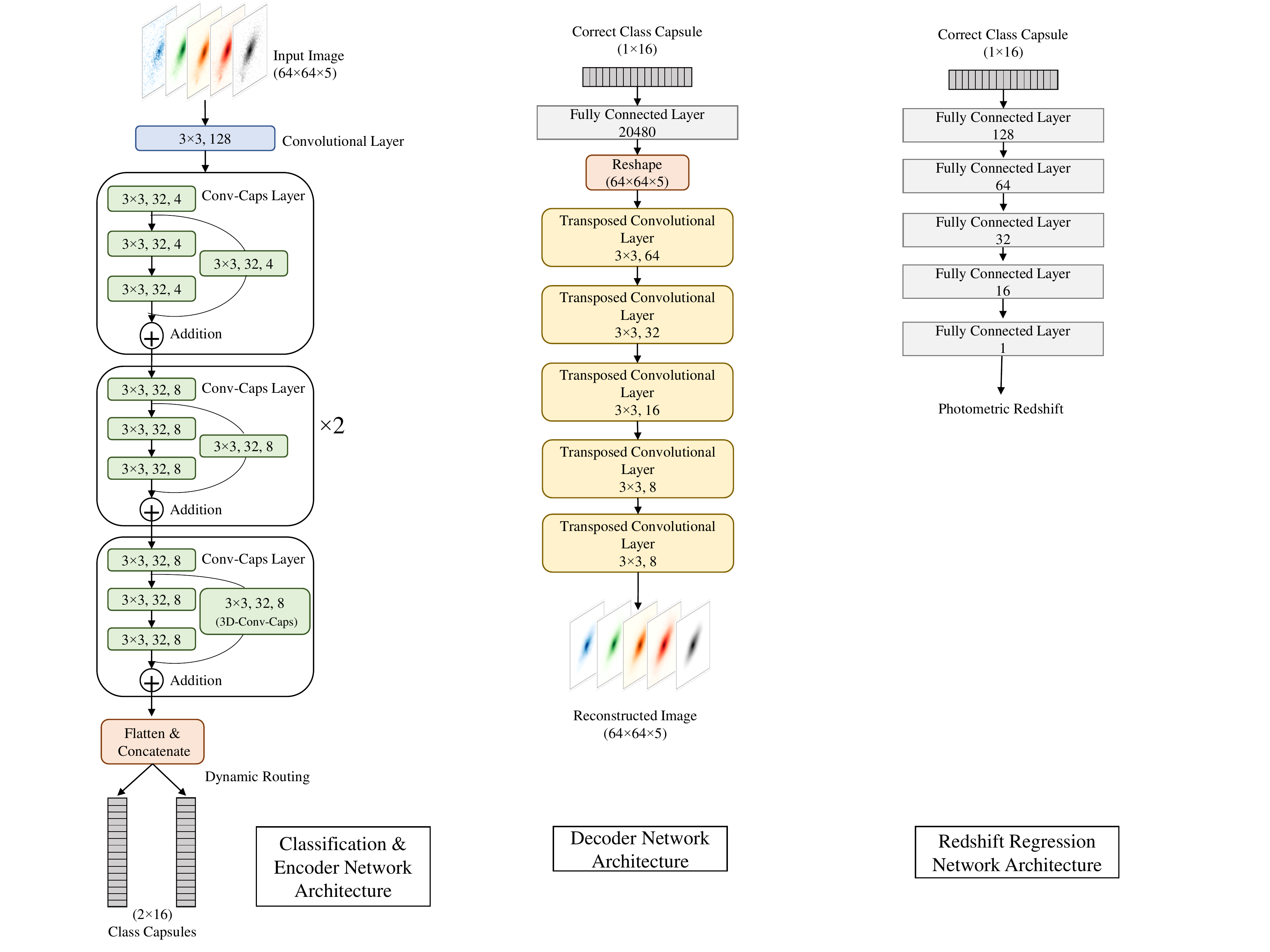}
  \caption{Schematic representation of the neural network architecture we use. The design of the classification and encoder network is based on \citet{RajasegaranEtal2019Deepcaps}. The classification network takes $ugriz$ images as inputs and produces two 16-dimensional capsule vectors as outputs, each representing a morphological class (spiral or elliptical). During training, the capsule corresponding to the correct morphological class is used as an input for the decoder and redshift regression networks whereas during inference the capsule vector with the largest magnitude (i.e., highest class probability) is used as the input for the subsequent networks. The numbers in each box represent the shape of the layer being used. For convolutional capsule layers (i.e. \texttt{Conv-Caps} and \texttt{3D-Conv-Caps} layers), they stand for the width $\times$ height, the number of capsules and the total number of dimensions for each capsule, respectively. For convolutional or transposed convolutional layers, they represent the width $\times$ height of the convolution filter kernel followed by the number of such filters being used. For fully connected layers, the number represents the number of nodes in the layer. We use a combination of the classification-and-encoding network and decoder network to generate morphological class labels for all the galaxies as a preliminary step and then use a combination of the three networks to predict redshifts. Details of the mathematical operations performed by the various kinds of capsule layers can be found in Appendix~\ref{appendix:routing}.}
  \label{fig:network-arch}
\end{figure*}

The network architecture we use has three main components: a deep capsule network-based classification-and-encoding network, a class independent decoder network, and a redshift prediction network. We use a combination of classification-and-encoding network and the decoder network to generate morphological class labels for the entire data set as a preliminary step and then use a combination of all three networks to jointly predict the morphology and photo-$z$ as described below and shown in Fig.~\ref{fig:network-arch}.

\paragraph*{The classification-and-encoding network} (Fig.~\ref{fig:network-arch} left column) inherits its architecture from \citet{RajasegaranEtal2019Deepcaps}. It takes the 5 band $64\times 64$ pixel images of a galaxy as inputs and uses a set of convolutional filters to convert the image into capsules. Next four blocks of skip connected convolutional capsule cells are used. The convolutional capsule layers were introduced in \citet{RajasegaranEtal2019Deepcaps} and use 3D-convolution operations to perform routing between two capsule layers more efficiently. Skip connections refer to the element-wise summing of outputs of an earlier layer with the output of a nonconsecutive layer ahead of it. This improves the convergence of the training process by preventing the gradients from vanishing and allowing information from earlier capsules to flow efficiently to later ones. The output of the final layer is a set of two 16-dimensional capsule vectors that we use to represent the spiral or elliptical morphological class of a galaxy. The Euclidean lengths of these capsules denote the probability of the input image being a spiral or elliptical. The individual dimensions of the vectors encode information about the input image, which can be used to predict the photometric redshift and reconstruct the input image. This part of the network has about 7.5 million trainable weights. 

\paragraph*{The class independent decoder network} (Fig.~\ref{fig:network-arch} middle column) is composed of successive transposed convolutional layers (also called de-convolution layers) which take one of the capsule vectors as input and try to reconstruct the input image as its output. Transposed convolution layers are mathematically similar to convolution layers except their input and outputs are switched. During the training process, we use the capsule representing the correct morphological class as the input of this network. During inference, the capsule with the largest length (i.e., the capsule representing the most probable class) is passed as the input to the decoder network. The decoder network acts as a regulariser and enforces that each dimension of the capsule vector represents a low-dimensional encoding of the input. The decoder network also helps us visually interpret the features encoded by the capsules. Using the same decoder network for both capsules (i.e., class independent decoder) makes the dimensions for both capsules represent similar properties. The decoder network has 0.88 million trainable weights. Some examples of the input and reconstructed images of galaxies are shown in Fig.~\ref{fig:recon}.

\paragraph*{The redshift regression network} (Fig.~\ref{fig:network-arch} right column) is a set of five fully connected neural network layers for redshift estimation. It takes as input the capsule corresponding to the correct morphological class during training and the capsule with the highest class probability during inference. This network has about 13,000 trainable weights.

\subsection{Loss Functions}
The weights of the networks are obtained by minimising a composite loss function which is a weighted sum of the losses calculated from the outputs of the three networks. The outputs from each of the networks are used to calculate a different loss function, a weighted sum of which is minimised depending on the task we are trying to solve. Following \citet{SabourEtal2017Capsnet}, we use the output of the classification-and-encoding network to calculate the Margin Loss (also called the Hinge loss) defined as:
\begin{equation}
 L_{\mathrm{margin}} =  \sum_{j=1}^{2} T_{j}\ \mathrm{max}(0, m^{+} - \lVert \mathbf{v}_{j} \rVert )^{2}) + \lambda (1-T_{j})\ \mathrm{max}(0, \lVert \mathbf{v}_{j} \rVert - m^{-})^{2}, \label{eq:margin-loss}
 \end{equation}
 where $T_{j}$ represent the class labels and $T_{j}=1$ when a galaxy corresponding to class $j$ is present in the input image and $T_{j}=0$ otherwise, $m^{+}=0.9$, $m^{-} = 0.1$ and $\lambda=0.5$. The parameters $m^{+/-}$ define a threshold for the length of the capsule above which the classification is considered correct/incorrect. The $\lambda$ parameter down-weights the margin loss for an absent morphological class, preventing the lengths of all the capsules from shrinking during the initial learning phase. The loss is summed over each class (2 in our case). This loss function is optimised to ensure that the length of one of the capsules is close to 1 and the other one close to 0 when the input is a spiral galaxy and vice versa when the input is an elliptical galaxy.
 
 We use the output of the decoder network to calculate the sum of squared errors between the input and reconstructed image pixels defined as:
 \begin{equation}
     L_{\mathrm{decoder}} = \sum_{k=1}^{5}\sum_{j=1}^{64}\sum_{i=1}^{64}(x_{ijk}-\hat{x}_{ijk})^{2},
 \end{equation}
 where, $\mathbf{x}$ and $\mathbf{\hat{x}}$ denote the input and reconstructed images respectively and the summation is carried out over all the $64\times64$ pixels and $5$ imaging bands.
 
 Similarly, we use the output of the redshift regression network to calculate the squared error between the spectroscopic redshift and the predicted photometric redshift defined as:
 \begin{equation}
     L_{\mathrm{photo-}z} = (z_{\mathrm{spec}}-z_{\mathrm{phot}})^{2},
 \end{equation}

All the losses are then averaged over the number of objects present in the training batch. The exact weighting of these losses will be discussed in the next two sections.
 
 \section{Training Procedure} \label{sec:training}
 \subsection{Generating Morphological Class Labels}\label{sec:morphology-classification}
 Morphological class labels from Galaxy Zoo-1 are available for only 34\% of the galaxies in our data set (see Sec.~\ref{sec:data-morpho}). We follow a fully supervised learning approach, and our capsule network design relies on the availability of morphological class labels. Therefore, we need to generate morphological class labels for the remainder of the data set to train the network to predict redshifts. To achieve this, we train a deep capsule network that is a combination of the classification-and-encoding network and the decoder network. The decoder network acts as a regulariser. We minimise the weighted sum of the margin loss for classification and the total squared error for reconstruction with a weight of 1 on the margin loss and $0.005$ on the reconstruction loss. So, for this task we the loss function ($L$) given by:
 \begin{equation}
     L = L_{\mathrm{margin}} + 0.005\times L_{\mathrm{decoder}},
 \end{equation}

 We divide the set of 177,442 galaxies with good morphological class labels into a training set (80\%), validation set (10\%) and test set (10\%). We train the network to classify the galaxies as spirals or ellipticals and achieve over 99\% classification accuracy on the test set. We then use this network to predict morphology labels for the galaxies that do not have a label from Galaxy Zoo-1. We then calibrate the predicted class probabilities with isotonic regression \citep{Zadrozny2001IsotonicCallib, Zadrozny2002IsotonicCallib} using the validation set for training the isotonic regression model and the test set to verify the calibration. This step ensures that the class probabilities predicted by the network are statistically consistent. We then select galaxies with calibrated class probabilities over 0.8, assign them to their corresponding class label and merge them with the initial training set. We train the same network again with this new training set and follow the same procedure above to assign labels and extend the training set. We do this step one more time and find that 99.6\% of the galaxies in our parent set has a class label with more than 0.8 class probability. For the remaining 0.4\% of the galaxies, we assign a label corresponding to the class with the highest probability. 
 
 We are generating morphological class labels for 339,083 galaxies based on a human labelled training set of 177,442 galaxies. The bulk of the galaxies do not have a confident morphological class label in Galaxy Zoo-1 as a strong consensus was not achieved among the human volunteers. This either might be because the shape of the galaxy is ambiguous or there were some artefacts in the image. A visual inspection of the galaxies in the test set which do not have a label from Galaxy Zoo-1 shows that the objects can almost always be classified into a spiral or elliptical galaxy by the authors and the predictions of our model for those objects matches with the judgement of the authors. The number of images which have ambiguous morphology or where an artefact or merger makes the morphology difficult to infer are negligibly small (~0.1\%). Since our main goal is to improve photo-$z$ prediction performance, we are comfortable with using the smaller training set with only good classifications to generate class labels for the entire data set and ignoring the very small number of ambiguous cases. As a separate cross-check we compared the class labels generated by our method for the galaxies which do not have a confident label from Galaxy Zoo-1 with the most voted Galaxy Zoo-1 class label and find that they are in agreement for over 70\% of the objects.

\subsection{Training for Photo-$z$ Estimation}
Once we have morphological class labels for all the galaxies in our data set, we  now train a neural network that is a combination of the classification-and-encoding network, the redshift regression network, and the class independent decoder network. The classification-and-encoding network gives us a low-dimensional representation of the input image which is then used by the redshift regression network to predict the photometric redshift. Although the decoder network doesn't directly help with redshift prediction, it has been shown to have a regularisation effect on capsule networks \citep{SabourEtal2017Capsnet}. The decoder network also ensures that the low-dimensional encoding learnt has physically meaningful information, which can be used to reconstruct the input image. In Sec.~\ref{sec:tinker}, we use the decoder network to interpret the features learnt by the capsule network.

\begin{figure*}
  \begin{subfigure}[b]{0.35\textwidth}
    \includegraphics[width=\textwidth]{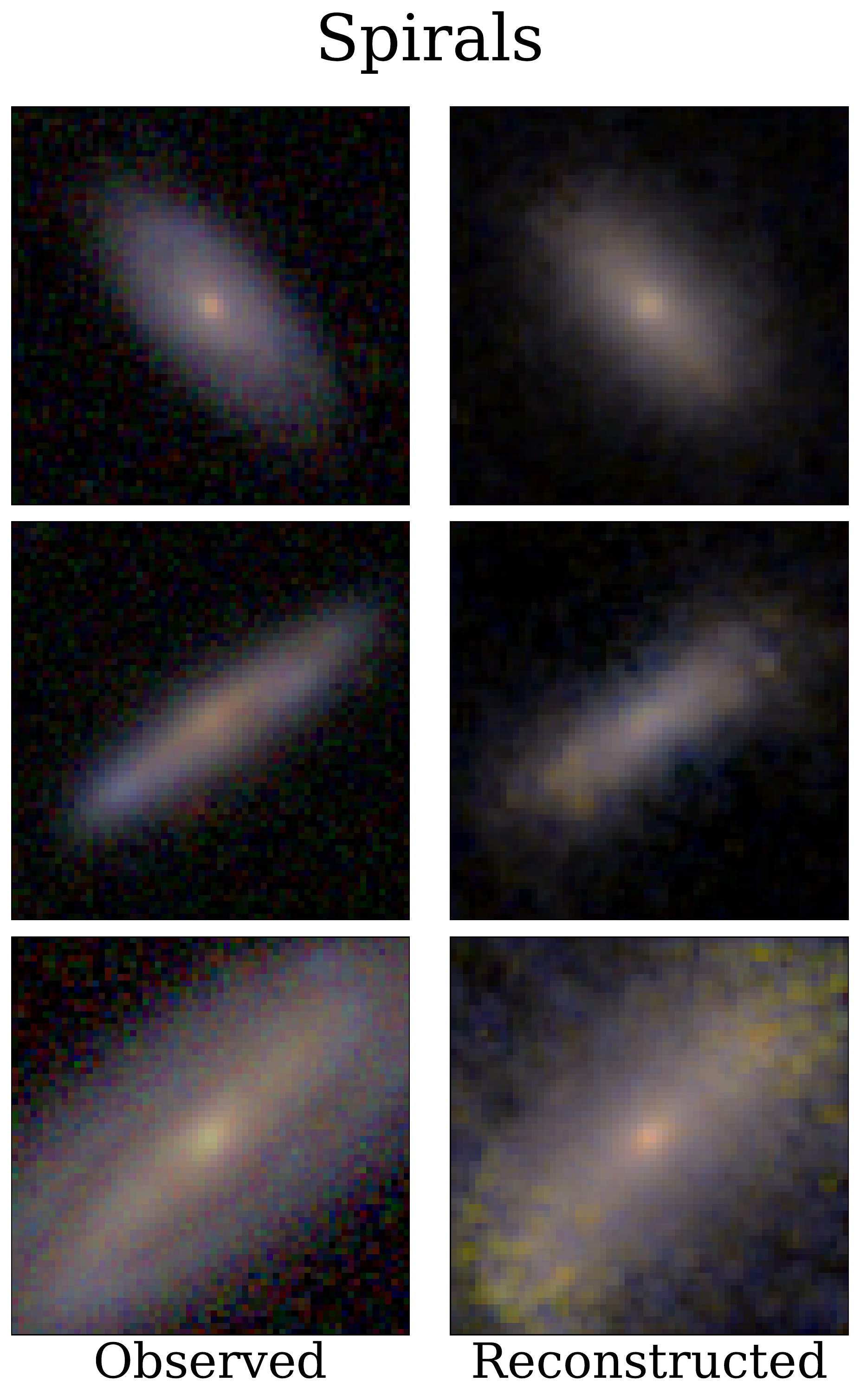}
    \label{fig:disks}
  \end{subfigure}
\hskip 4cm
  \begin{subfigure}[b]{0.35\textwidth}
    \includegraphics[width=\textwidth]{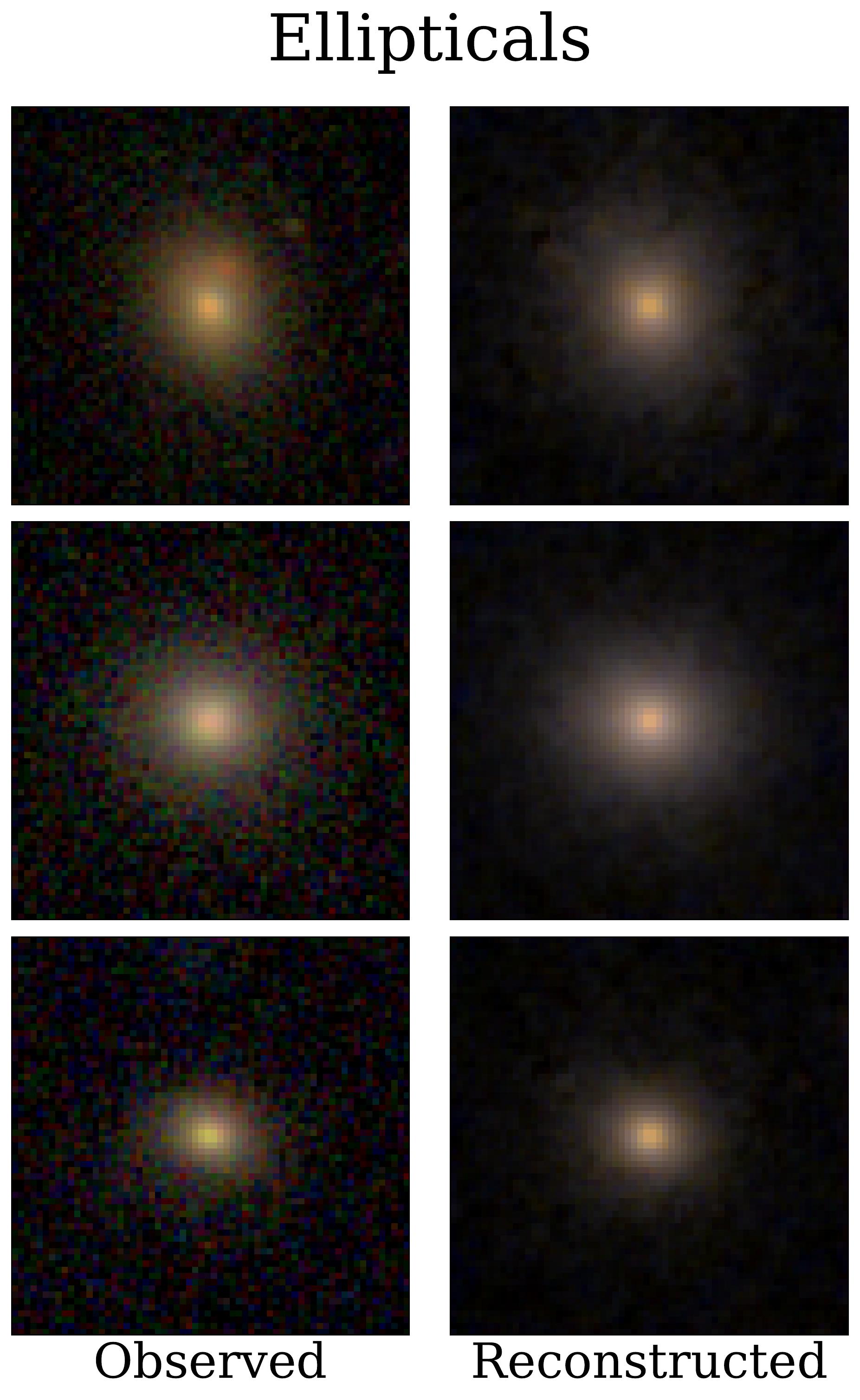}
    \label{fig:spheroids}
  \end{subfigure}
\caption{Comparison of the observed and reconstructed $grz$ images of a few randomly selected spirals (left) and ellipticals (right) from the test set. The reconstructions were produced by the decoder network using the 16-dimensional capsule corresponding to the predicted morphological type. We see that the reconstructions capture basic properties of the input like shape, orientation, and colour.}
\label{fig:recon}
\end{figure*}

During the training process, the capsule corresponding to the correct morphological class is used as an input for both the decoder and redshift regression networks whereas during inference the capsule vector with the largest Euclidean length (i.e., highest class probability) is used as their inputs. To find the optimum set of weights for the network, we minimise a composite loss function which is a weighted sum of the losses from each of the three networks. Similar to Sec.~\ref{sec:morphology-classification}, we use the weighted sum of the margin loss and total squared error for the classification and reconstruction tasks, but now we also add the squared error of the predicted redshift to the total loss ($L$):
\begin{equation}
     L = L_{\mathrm{margin}} + 0.005\times L_{\mathrm{decoder}} + L_{\mathrm{phot-}z}.
\end{equation}
The classification, reconstruction, and redshift regression losses are given the weights of 1, 0.005, and 1 so that they contribute an equal amount towards the total value of the loss. This allows us to put equal importance on each of the individual tasks as all of them help to improve the accuracy of photometric redshifts. Some examples of reconstructed images of galaxies obtained after training the network are shown in Fig.~\ref{fig:recon}.

Instead of directly predicting the redshifts, we scale the redshifts using the logistic transformation defined as:
\begin{equation}
    h(z) = \log \left( \dfrac{z-z_{\mathrm{min}}}{z_{\mathrm{max}}-z}\right).
\end{equation}
For our data set $z_{\mathrm{min}}=0$ and $z_{\mathrm{max}}=0.4$. We find that performing this transformation gives us better performance especially at very low redshifts ($z<0.05$). This is because the logistic transformation makes the distribution of the target variable (redshift in our case) fall gradually at the boundaries, thereby alleviating the problem of attenuation bias.

We randomly split our data into three subsets; the training set which is used to train the network, the validation set, which is used to tune the hyper-parameters of the network and decide when to stop training and a test set which is used to check the final performance. All results quoted in this work use a training set that is 80\% the size of the parent data set and have been calculated on the test set which is 10\% the size of the parent set (unless stated otherwise). The remaining 10\% of the data is used as the validation set. We also check the performance of our photo-$z$ prediction as a function of the size of the training set as shown in Fig.~\ref{fig:performance_v_data}. 

To randomly initialise the weights for the networks, we use the He-Normal initialiser \citep{He2015HeNormalPrelu}. We use the PReLU \citep{He2015HeNormalPrelu} activation function for all the hidden layers and a linear activation function for the output layers of the decoder and redshift regression networks. To train all the networks, we use the Adam optimiser \citep{Kingma2015Adam} with an initial learning rate of 0.001. After each epoch the learning rate is decreased following the rule: $\mathrm{learning\ rate} = \mathrm{initial\ learning\ rate} \times 0.95^{\mathrm{epoch}}$. We also augment the training set by randomly rotating the images in steps of $90\degree$ or flipping them along the horizontal or vertical axis before passing them to the networks for training. The same setup is used for both the morphological label prediction and redshift estimation tasks.

We train the networks for 100 epochs but the training generally converges within 70 epochs. We choose the epoch which has the best performance---i.e., the highest classification accuracy when generating morphology labels and the lowest average redshift prediction error on the validation set. Since the model is initialised randomly, each training run can result in a different set of optimal weights. Hence we run the training process 5 times and take the average of their output as our photo-$z$ prediction. For this reason, we also select epochs that have a low bias and moderate variance since bias stays roughly the same whereas variance decreases when averaged.

The models are defined in Keras with Tensorflow 1.15 as the back end. The training is done on an Alienware Area 51 PC with an Intel Core i7 9800X processor, 2 RTX 2080Ti GPUs and 64GB of RAM. We use a batch size of 400 which takes about 8 hours to train for 100 epochs. The model is copied onto the two GPUs and the training is parallelised by sending half of the batch to each GPU.

\section{Results} \label{sec:results}
\subsection{Photo-$z$ Evaluation Metrics} 
\label{sec:metrics}
In this work, we are focusing only on photo-$z$ point estimates and not full PDFs. We will therefore assess the performance of our photo-$z$ estimates by measuring how much the spectroscopic and photometric redshifts for each galaxy in the test set differ. We use the following three common metrics:
\begin{itemize}
    \item  \textbf{Prediction bias} defined as $\langle \frac{\Delta z}{1+z_{\mathrm{spec}}} \rangle$, i.e. the average value of the prediction error.
    \item \textbf{Normalised Median Absolute Deviation ($\sigma_{NMAD}$)} defined as $1.4826 \times \mathrm{Median}(\mid \frac{\Delta z}{1+z_{\mathrm{spec}}} - \mathrm{Median}(\frac{\Delta z}{1+z_{\mathrm{spec}}}) \mid )$. This is a robust measure of the spread of prediction errors.
    \item \textbf{Fraction of Outliers (f$_{\mathrm{outlier}}$}) defined as the fraction of photo-$z$ predictions for which $\mid \frac{\Delta z}{1+z_{\mathrm{spec}}}\mid > 0.05$, i.e. the fraction of cases where the prediction error is very high. We chose the threshold of $0.05$ to easily compare our results with other similar works.
    
    The specific choice of the metrics and the threshold to define an outlier is based on convention and allows us to easily compare our results with recent similar work. 
\end{itemize}

\subsection{Photo-$z$ Point Estimate Predictions} \label{sec:point-estimates}

When trained on 80\% and tested on 10\% (with the remaining 10\% used as validation set) of the parent data set and results averaged over an ensemble of 5 models, our photo-$z$ estimates have $\sigma_{\mathrm{NMAD}}=0.00898$, f$_{\mathrm{outlier}}=0.19\%$ and $\langle \frac{\Delta z}{1+z_{\mathrm{spec}}} \rangle = 7\times10^{-5}$. For comparison, other deep learning based methods which take images as inputs like \cite{PasquestEtal2019Photoz} achieve $\sigma_{\mathrm{NMAD}}=0.00912$, f$_{\mathrm{outlier}}=0.31\%$ and $\langle \frac{\Delta z}{1+z_{\mathrm{spec}}} \rangle = 1\times10^{-4}$ when trained on the same data set and \citet{Hayat2021PhotozSelfSupervised} achieves $\sigma_{\mathrm{NMAD}}=0.00825$, f$_{\mathrm{outlier}}=0.21\%$ and $\langle \frac{\Delta z}{1+z_{\mathrm{spec}}} \rangle = 1\times10^{-4}$, by first pre-training on a large unlabelled data set (about twice as big as our data set) and then fine tuning on a data set similar to ours. Both of them use models with about 3 times as many trainable parameters compared to ours ($\sim$24 million vs $\sim$8 million). Our algorithm has comparable $\sigma_{\mathrm{NMAD}}$ and better f$_{\mathrm{outlier}}$ performance among these deep learning based methods.

\begin{figure}
 \includegraphics[width=\linewidth]{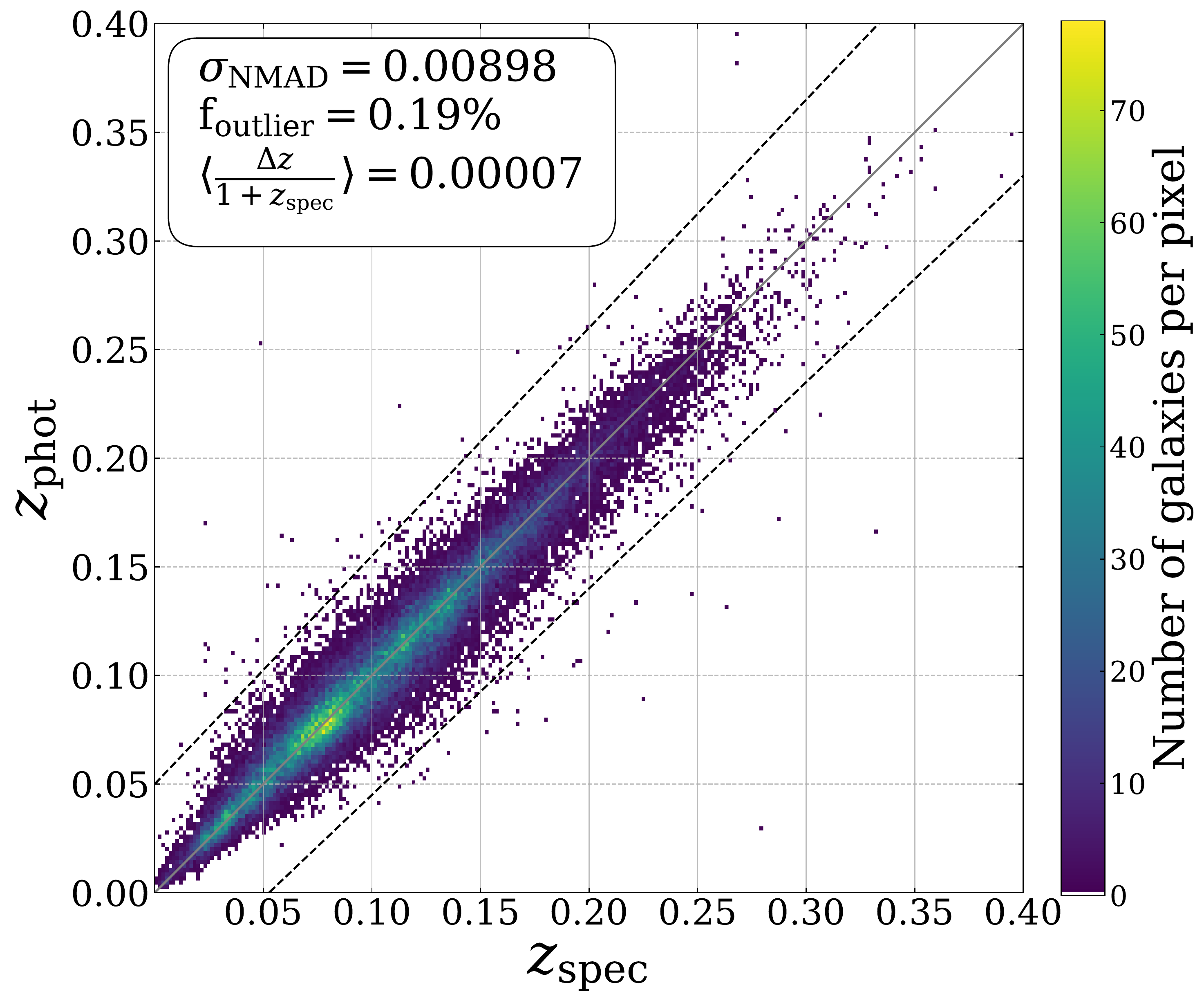}
 \caption{Comparison of photometric redshift point estimates predicted by our capsule network with the corresponding spectroscopic redshifts for galaxies in the test set. The central grey line shows $z_{\mathrm{phot}}=z_{\mathrm{spec}}$, i.e., a perfect photo-$z$ estimate. The outer dashed lines mark $\mid \frac{\Delta z}{1+z_{\mathrm{spec}}}\mid = 0.05$. Any point lying outside these limits (i.e. $\mid \frac{\Delta z}{1+z_{\mathrm{spec}}}\mid > 0.05$ ) is considered to be an outlier. The colour on the scatter plot shows the number of data points present in each pixel of the figure. We see that the scatter is tight and symmetrically distributed about the $z_{\mathrm{phot}}=z_{\mathrm{spec}}$ line and with a negligible bias. The scatter looks random and shows no visible patterns at the limits of training data ($z_{\mathrm{spec}}\approx0$ and $z_{\mathrm{spec}}>0.3$) indicating stable performance across the redshift range. }
 \label{fig:point-est}
\end{figure}

We show a comparison between the photometric and the spectroscopic redshifts for the test set in Fig.~\ref{fig:point-est}. We see that the scatter is tight and distributed symmetrically about the $z_{\mathrm{phot}}=z_{\mathrm{spec}}$ line. The scatter in the points and distribution of outliers look random and show no visible patterns of a sudden change in performance at the limits of training data ($z_{\mathrm{spec}}\approx0$ and $z_{\mathrm{spec}}>0.3$) indicating stable performance across the redshift range. We also see no evidence of attenuation bias (i.e., almost constant predictions for a subset of inputs; see \citealt{Freeman2009AttenuationBias} for a discussion on attenuation bias in photo-$z$ algorithms).
The images used to train the networks include observations of Stripe 82 \citep{Jiang2014Stripe82}, which are about 2 magnitudes deeper and have less noise than the rest of the images. Since Stripe 82 is a small fraction ($<4\%$) of the whole data set, we do not account for this varying depth by weighing data points differently. We find a significantly smaller spread in the predictions ($\sigma_{\mathrm{NMAD}}=0.00741$) and a fraction of outliers consistent with the rest of the sample (f$_{\mathrm{outlier}}=0.35\%$), given the small number of Stripe 82 objects in the test set. Galaxies in the test set outside of Stripe 82 produce photo-$z$'s with $\sigma_{\mathrm{NMAD}}=0.00906$ and f$_{\mathrm{outlier}}=0.19\%$. This shows that having images with a higher signal-to-noise ratio improves the quality of photo-$z$ predictions.

When the test set is split into subsets based on morphology, we find that the photo-$z$ predictions have a lower spread for ellipticals than spirals ($\sigma_{\mathrm{NMAD}}=0.00844$ vs. $0.00956$) with a comparable fraction of outliers (0.18\% vs 0.20 \%). This might be because elliptical galaxy populations have similar rest-frame colours as older stellar populations tend to change very little in colour with time. The observed colours and magnitudes (or any other measure of flux) therefore trace the redshift well making it is easier to predict redshifts of elliptical galaxies than spirals. When we split the test set based on the availability of human labelled morphology, we find that photo-$z$ prediction performance is better when human labelled morphology is available ($\sigma_{\mathrm{NMAD}}=0.00815$, f$_{\mathrm{outlier}}=0.11\%$ vs. $\sigma_{\mathrm{NMAD}}=0.00948$, f$_{\mathrm{outlier}}=0.23\%$). Although human labelled morphology improves the performance, the lack of it does not reduce the performance drastically.

We performed a visual inspection of the images of the galaxies which were photo-$z$ prediction outliers. We find that around $18\%$ of these outliers have bad or missing photometry. Removing these objects from our test set reduces our outlier fraction to f$_{\mathrm{outlier}}=0.16\%$. We kept these rare objects in the parent data set for easy comparisons with \citet{PasquestEtal2019Photoz}.

\begin{figure}
 \includegraphics[width=\linewidth]{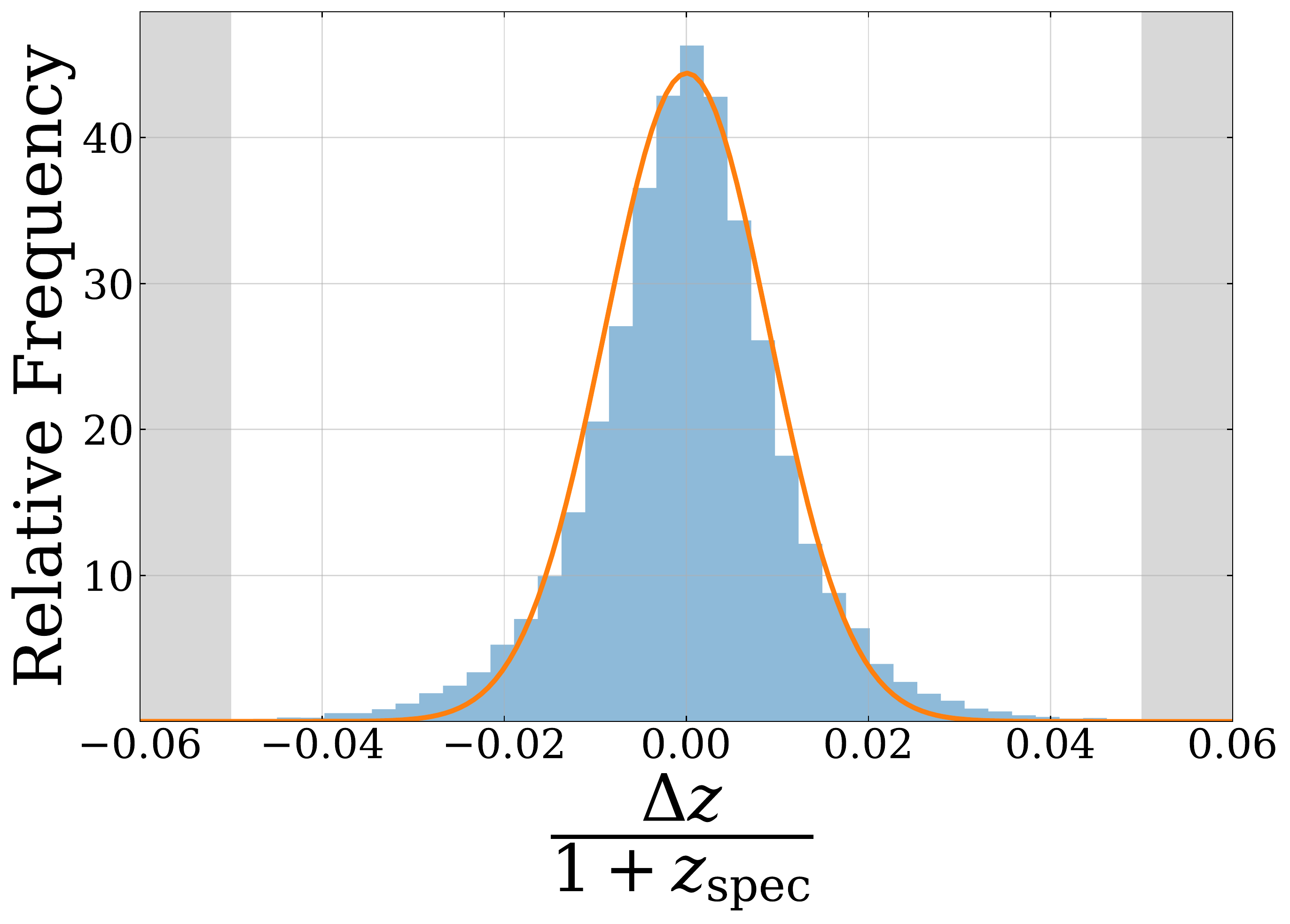}
 \caption{Normalised distribution of the redshift prediction errors. The blue histogram shows the distribution of redshift prediction errors of our algorithm on the test set. The orange line shows a Gaussian distribution with the location and scale parameters set as the prediction bias and $\sigma_{\mathrm{NMAD}}$ respectively. The distributions are normalised to have unit area under the curves. The shaded region marks the threshold for outliers. The distribution of the prediction errors is symmetric, centred around 0 and closely resembles a Gaussian distribution, indicating little if any systematic preference for over- or under-estimation.}
 \label{fig:error_dist}
\end{figure}

The distribution of prediction errors is shown in Fig.~\ref{fig:error_dist}. They follow a symmetric distribution centred about 0 indicating little if any systematic preference for over or under-estimation. Since the fraction of outliers is very small and we see that the distribution of prediction errors closely resembles a Gaussian distribution, $\sigma_{\mathrm{NMAD}}$ can be treated as the $1\sigma$ Gaussian uncertainty around each prediction up to a good approximation. 

\begin{figure*}
 \includegraphics[width=\linewidth]{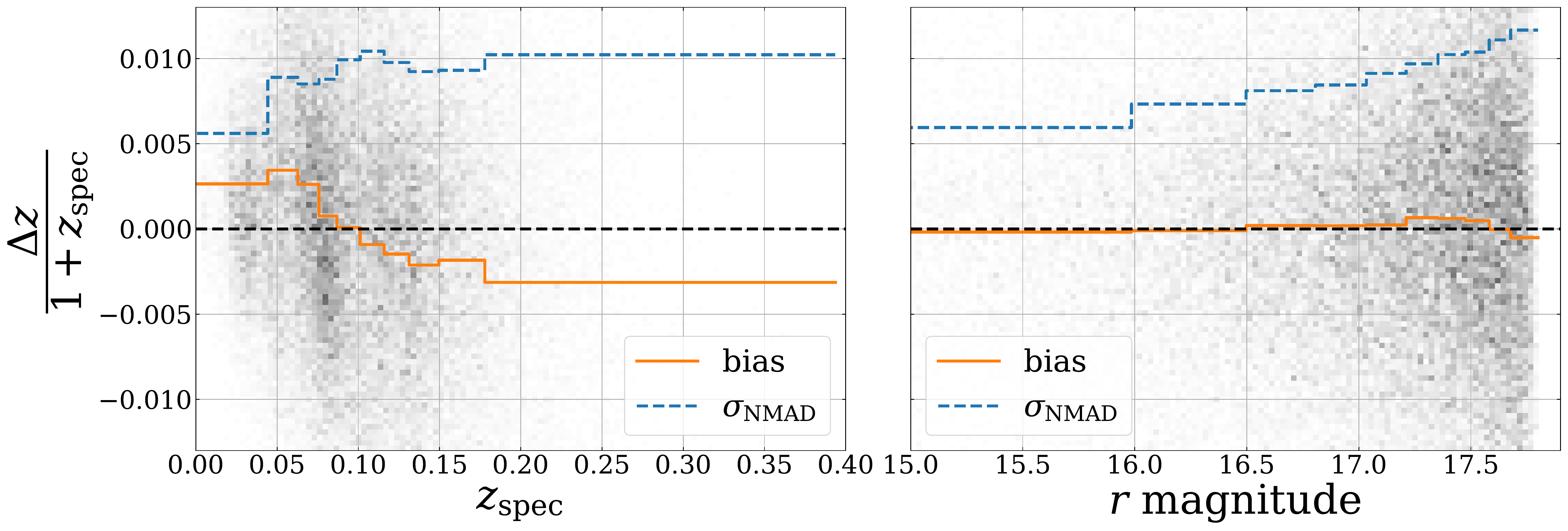}
 \caption{Prediction bias ($\langle \frac{\Delta z}{1+z_{\mathrm{spec}}} \rangle$) and $\sigma_{\mathrm{NMAD}}$ of our photometric redshift estimates as a function of spectroscopic redshift ($z_{\mathrm{spec}}$, left) and $r$-band Petrosian magnitude (right). The metrics have been calculated for 10 bins of equal population. We use bins with varying widths but equal populations so that the standard errors on the binned statistics are comparable across all bins. The grey points show the distribution of individual galaxies. Due to the relatively large number of samples in each bin, the standard errors on the statistics are very small. We see that $\sigma_{\mathrm{NMAD}}$ increases at higher redshifts (where we have less training data) and for fainter galaxies (where the signal to noise ratio of the images are lower). Though the bias on average is very small, it is higher at the lowest and highest redshift bins but with opposite signs with an inflection at the median $z_{\mathrm{spec}}$ ($\approx0.1$). The bias is constant and negligible in magnitude over the entire range of $r$-band Petrosian magnitudes. }
 \label{fig:met_v_z}
\end{figure*}

We also check the performance (prediction bias and $\sigma_{\mathrm{NMAD}}$) of our photo-$z$ estimates as a function of the spectroscopic redshift and $r$-band Petrosian magnitude of galaxies as shown in Fig.~\ref{fig:met_v_z}. We use the Petrosian magnitude as it was used to define the faintness cut of the data set we are using. As a function of redshift, the absolute magnitude of the bias is small though it is positive at low redshifts and negative at high redshifts with the inflection point being at the median redshift ($\approx0.1$) of our data set. This kind of pattern is common for ML-based algorithms. When seen as a function of $r$-band magnitude, the bias is almost constant and negligibly small in magnitude throughout the entire range of magnitudes. $\sigma_{\mathrm{NMAD}}$ tends to increase both as we go to higher redshifts and fainter magnitudes. This can be attributed to the fact that there is less training data and increased noise in the images at these regimes. We also see that $\sigma_{\mathrm{NMAD}}$ ($\sim0.006$) is significantly lower than the global value at low redshifts ($z<0.05$) even though the number of training samples available is small in this regime due to lower survey volume. We suspect this is because, at very low redshifts resolved information in the images like morphology, size, and surface brightness. contains rich information about galaxy distances. Better photo-$z$ performance at very low redshifts can aid in the identification of satellite galaxies that require a massive spectroscopic effort to get redshifts (e.g., \citealt{Geha2017SAGA}, \citealt{Mao2021SAGA}).

\begin{figure*}
  \includegraphics[width=1\linewidth]{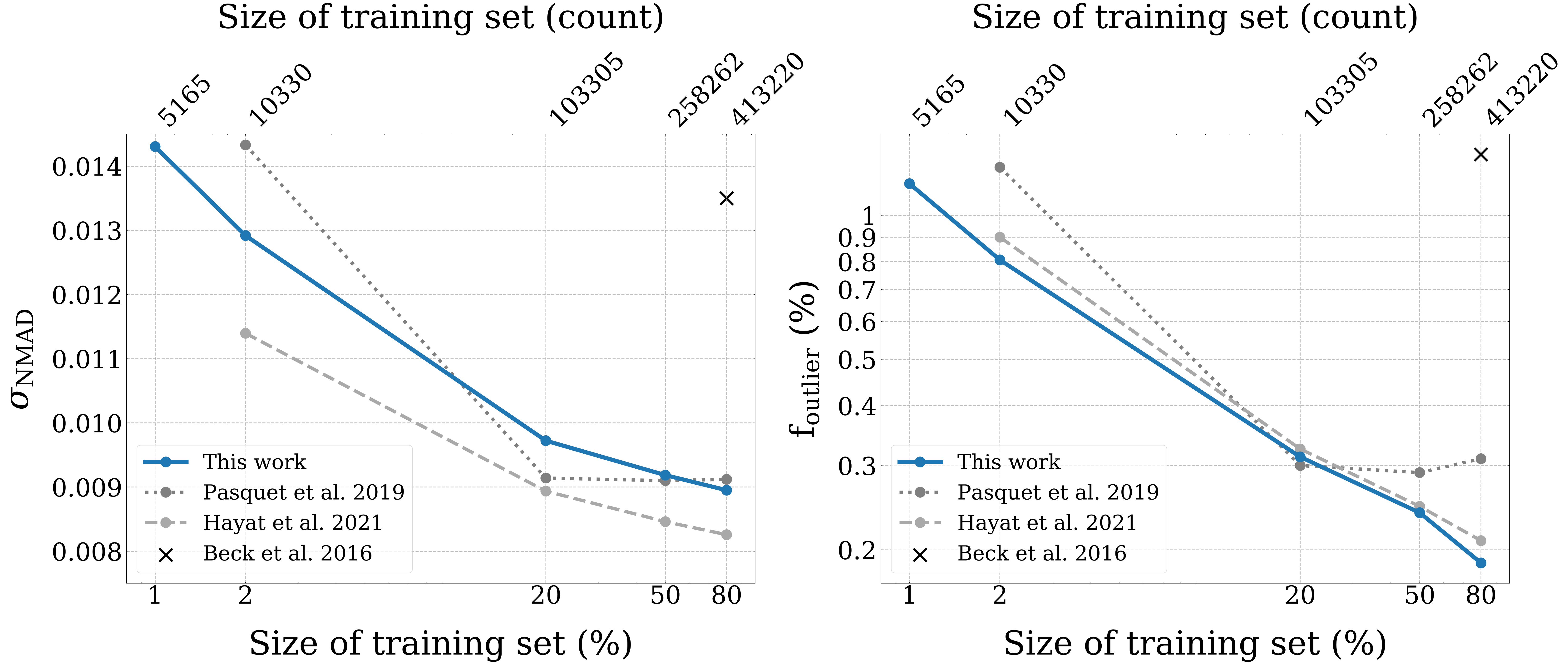}
  \caption{Performance of photometric redshift prediction algorithm as a function of the size of training data. The standard errors on the statistics are negligibly small and hence not shown. Our algorithm has comparable $\sigma_{\mathrm{NMAD}}$ and better  $\mathrm{f}_{\mathrm{outlier}}$ performance to the two deep learning-based efforts \citep{PasquestEtal2019Photoz,Hayat2021PhotozSelfSupervised} and significantly better performance than the classical ML-based technique \citep{Beck2016Photoz} while requiring less training (or pre-training) data and fewer trainable parameters ($\sim$8 million vs. $\sim$23 million). }
  \label{fig:performance_v_data}
\end{figure*}

Obtaining spectroscopic redshifts is often an expensive process, so it is important that machine learning-based methods can perform well when the training data sets are smaller. To see how the photo-$z$ performance of our algorithm changes, we train our capsule network-based model using varying sizes of training data by random sub-sampling of the parent data set (after obtaining morphological labels) into smaller subsets while keeping everything else the same in the training process. The results are shown in Fig.~\ref{fig:performance_v_data} and also compared with other similar works like \citet{PasquestEtal2019Photoz}, \citet{Hayat2021PhotozSelfSupervised} and \citet{Beck2016Photoz}. The data for \citet{PasquestEtal2019Photoz} and \citet{Beck2016Photoz} were obtained from Table 2 in \citet{PasquestEtal2019Photoz}, the data for \citet{Hayat2021PhotozSelfSupervised} were obtained from their Fig.~4 using the WebPlotDigitizer \citep{Rohatgi2020WebplotDigitizer}. The metrics for \citet{Beck2016Photoz} provided here are calculated on their photo-$z$ estimates of the same objects as ours. They train on a much larger data set spread over a larger redshift range compared to ours which maybe one reason for higher prediction errors. We always use $10\%$ of the parent data set as the validation set and use the remaining amount of data to test the performance. We observe that we outperform \citet{Beck2016Photoz}, which is a widely used source of SDSS photo-$z$ estimates using just 2\% of the parent sample (or $\sim 10^4$ galaxies) as a training set. Many surveys of the high redshift Universe like CANDELS \citep{GroginEtal2011CandelsSurvey,Koekemoer2011CandelsData} have spectroscopic observations for a similar number of galaxies, albeit across a larger redshift range and our method could potentially be used to improve the photo-$z$ estimates for them. We see that our method has performance comparable to other deep learning-based photo-$z$ estimation methods like \citet{PasquestEtal2019Photoz} or \citet{Hayat2021PhotozSelfSupervised} when both are trained on random subsets of data. 

\subsection{Interpreting the Features Learnt by the Capsule Network }
As machine learning-based methods have started replacing more traditional physics-based methods to model astrophysical phenomena and make predictions that reduce the need for making extra observations, it is becoming increasingly important to peer inside these complex mathematical models to identify what physical features they are learning. This will not only help us to validate what the algorithms are predicting but also help us bridge the gap between the traditional physics-driven and the newer data-driven approaches. 

In our work, we use the capsule vectors along with the decoder network to shed some light on the features learnt by the network. Since the capsules composing the output layer of the morphology classification network are trained to represent a morphological class of galaxies along with holding enough information to predict the redshift and a reconstruction of the input image, we expect the components of the capsule vector to learn a low-dimensional encoding of the input galaxy image. Moreover, we expect that each of the component will learn properties so that all capsule dimensions combined can effectively predict the morphology, redshift and a reconstruction of the input image. 

The features learnt by the networks are not constrained to be easily identifiable visual properties or commonly used physical quantities derived from images. We will therefore perform both visual exploration of these features and also measure how well these features correlate with galaxy properties.

\subsubsection{Visualising the Capsule Encoded Space}
We first take a look at how the capsules corresponding to each galaxy in the test set are organised in their manifold. We use Uniform Manifold Approximation and Projection (UMAP; \citealt{McInnes2018UMAP}) to embed the 16-dimensional capsules into a 2-dimensional space to visualise and interpret any structures, if present. UMAP is a non-linear dimensionality reduction method that uses techniques from manifold learning  and topological data analysis to embed a high-dimensional data set into a low-dimensional manifold. To ensure that the relative local density of data is preserved when we project the capsules onto a 2-dimensional space, we use DensMAP \citep{Narayan2020DensMAP}, which computes the estimates of local density and uses them as a regulariser in the optimisation of the 2D UMAP representation. UMAP with the DensMAP regulariser preserves the local structure of the data while capturing global structure better than many other similar algorithms and is also computationally efficient. 

\begin{figure*}
 \includegraphics[width=\linewidth]{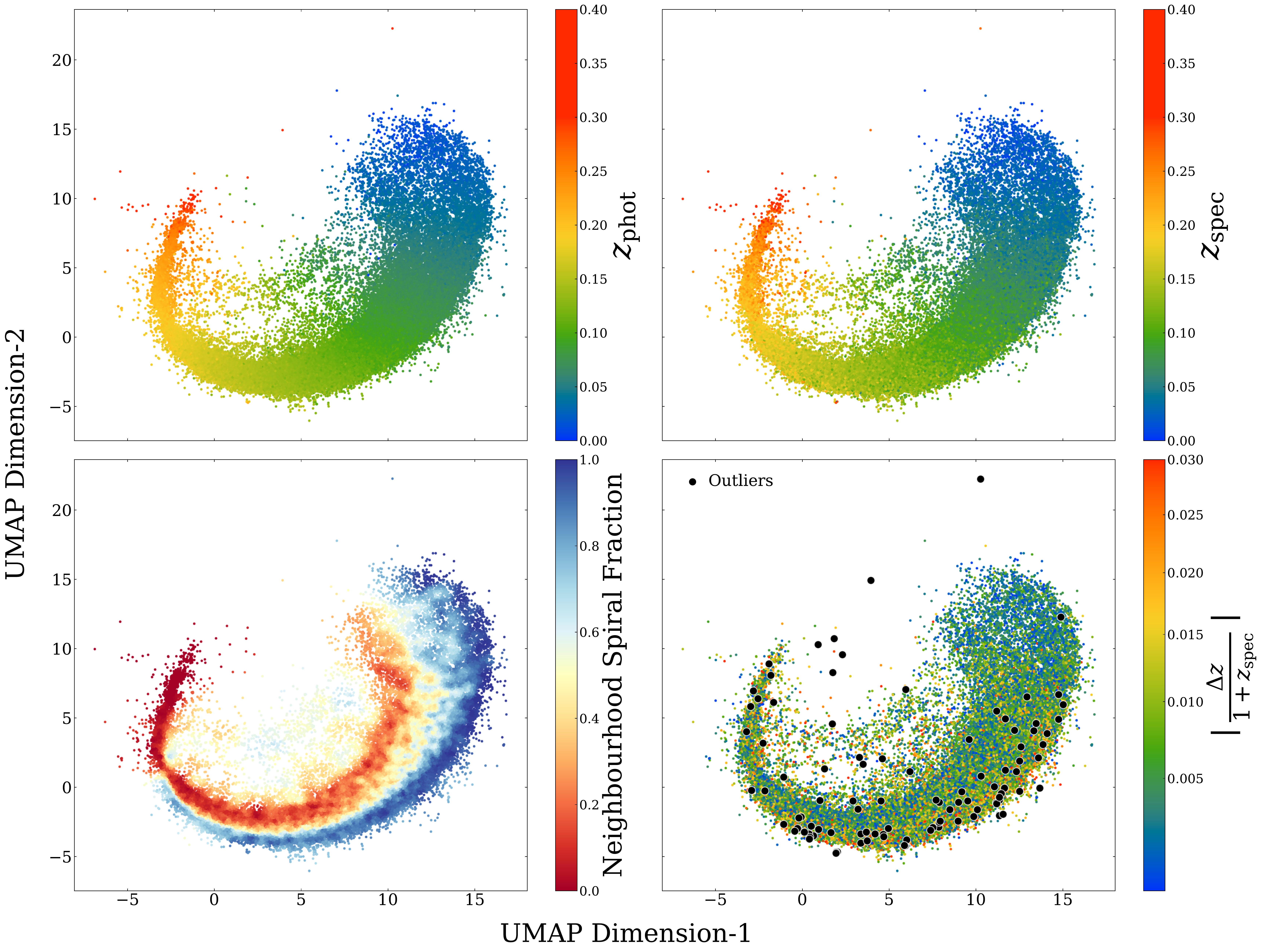}
 \caption{Two dimensional UMAP embedding of the 16-dimensional capsules colour-coded by photometric redshift (top left), spectroscopic redshift (top right), fraction of spiral galaxies in the neighbouring region (bottom left), and redshift prediction error (bottom right). The photo-$z$ outliers are shown in black in the bottom right panel. The UMAP embedding of the capsules creates a nearly perfect redshift sequence indicating that the capsules learn a good representation of redshift. Spirals and ellipticals tend to occupy separate regions though there is a region with overlap with morphology producing a gradient almost perpendicular to the redshift sequence. We notice that regions dominated by spirals tend to have slightly higher redshift prediction errors compared to regions dominated by ellipticals. Though spirals and ellipticals have almost the same fraction of outliers, visually it may seem that there are more outliers which are spirals than ellipticals. Many of those outliers are actually ellipticals which lie close to the region dominated by the spiral galaxies in the 2D UMAP representation. An interactive version of this figure showing galaxy image thumbnails is available online\protect\footnotemark.}
 \label{fig:umap}
\end{figure*}

Fig.~\ref{fig:umap} shows the 2-dimensional UMAP embedding of the 16-dimensional capsules colour coded by various properties. When coloured by photometric or spectroscopic redshift (top row), the embedding shows a nearly perfect redshift sequence. As UMAP places nearby capsules in the high-dimensional space close together in their 2-dimensional projection, we can infer that the capsules track a smooth redshift sequence. This is in contrast to the representations generated by self-organising maps (SOMs; \citealt{kohonen1981SOM,kohonen1982SOM}), which group galaxies with similar spectral energy distributions together using their photometry but impose a geometry that can force adjacent cells to have wildly different redshifts \citep{Masters2015SOMCallib}.  Currently, SOMs are widely used to determine regions with incomplete spectroscopic data (e.g., \citealt{Masters2015SOMCallib,Masters2019SOMCallib}), but dimensionality reduced capsules may perform better at this task due to its smooth redshift distribution.

If we colour the points based on the fraction of spirals among the 80 nearest neighbours in the 2D space (bottom left), we see that the spirals and ellipticals tend to occupy separate regions of the space although there is a significant overlap. The fraction of spirals exhibits a gradient almost perpendicular to the redshift sequence thereby effectively encoding both redshift and morphology, properties the capsules were trained to learn.  When colour-coded by the redshift prediction errors (bottom right) and compared with the plot showing the fraction of neighbouring spirals, we notice that regions dominated by spirals tend to have slightly higher redshift prediction errors compared to regions dominated by ellipticals. This was quantified in Sec.~\ref{sec:point-estimates} where we noted that spirals have slightly higher value of $\sigma_{\mathrm{NMAD}}$ compared to ellipticals but equivalent  $\mathrm{f}_{\mathrm{outlier}}$. Visually from Fig.~\ref{fig:umap} it may seem that there are more outliers which are spirals than ellipticals but many of those outliers are ellipticals which lie close to the region dominated by the spiral galaxies in the 2D UMAP representation.

Most of the galaxies in the 2D UMAP representation lie on the large crescent shaped sequence. A small number (about 1-2\%) of galaxies deviate from this sequence forming a smaller sequence encircled by the larger crescent. These galaxies all have higher values for dimension 10 of their capsules. Synthetic images generated by perturbing capsule dimensions (see appendix~\ref{appendix:tinker}) shows that higher values of dimension 10 tend to increase the extended component of the galactic disk. Some of the dimension 10 outliers in the main redshift sequence clearly have stars in the image. However, our investigation of the dimension 10 outliers in the smaller sequence has yet to yield a clear interpretation. These galaxies are a 50/50 mix of spirals and ellipticals, and the majority do not have neighbouring stars, galaxies, or artefacts.  A systematic study of these outlier galaxies will be done in a future work. The other galaxies that randomly scatter away from the two large sequences almost always have a neighbouring star, galaxy, or an artefact.

\label{sec:tinker}

\subsubsection{Generating Synthetic Images by Perturbing Capsule Dimensions } \label{sec:tinker}
To check whether the components of the capsules represent any visually identifiable properties of the galaxies, we take the capsule corresponding to the predicted morphology of a galaxy and add a small perturbation to one of the components keeping all the others fixed.  The perturbation is added in units of standard deviation of the values of the components in our test set. We pass on this perturbed capsule vector to the decoder network to see how the reconstructed image of the input changes. 

\begin{figure*}
    \begin{subfigure}[b]{1\textwidth}
    \centering
    \includegraphics[width=1\linewidth]{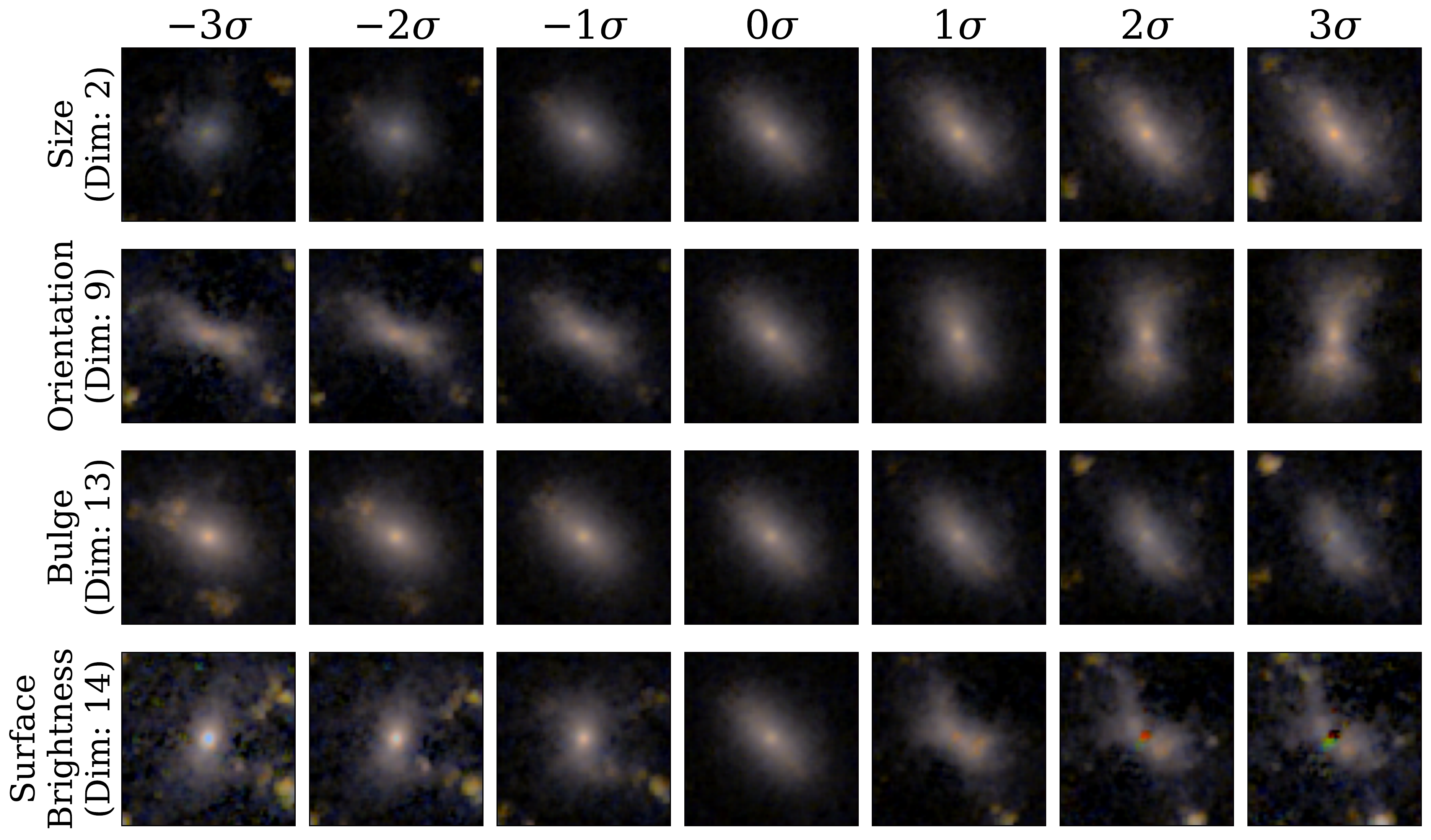}
    \caption{The first spiral galaxy from Fig.~\ref{fig:recon}}
    \label{fig:tinker_disk}
  \end{subfigure}
\newline
\begin{subfigure}[b]{1\textwidth}
\centering
  \includegraphics[width=1\linewidth]{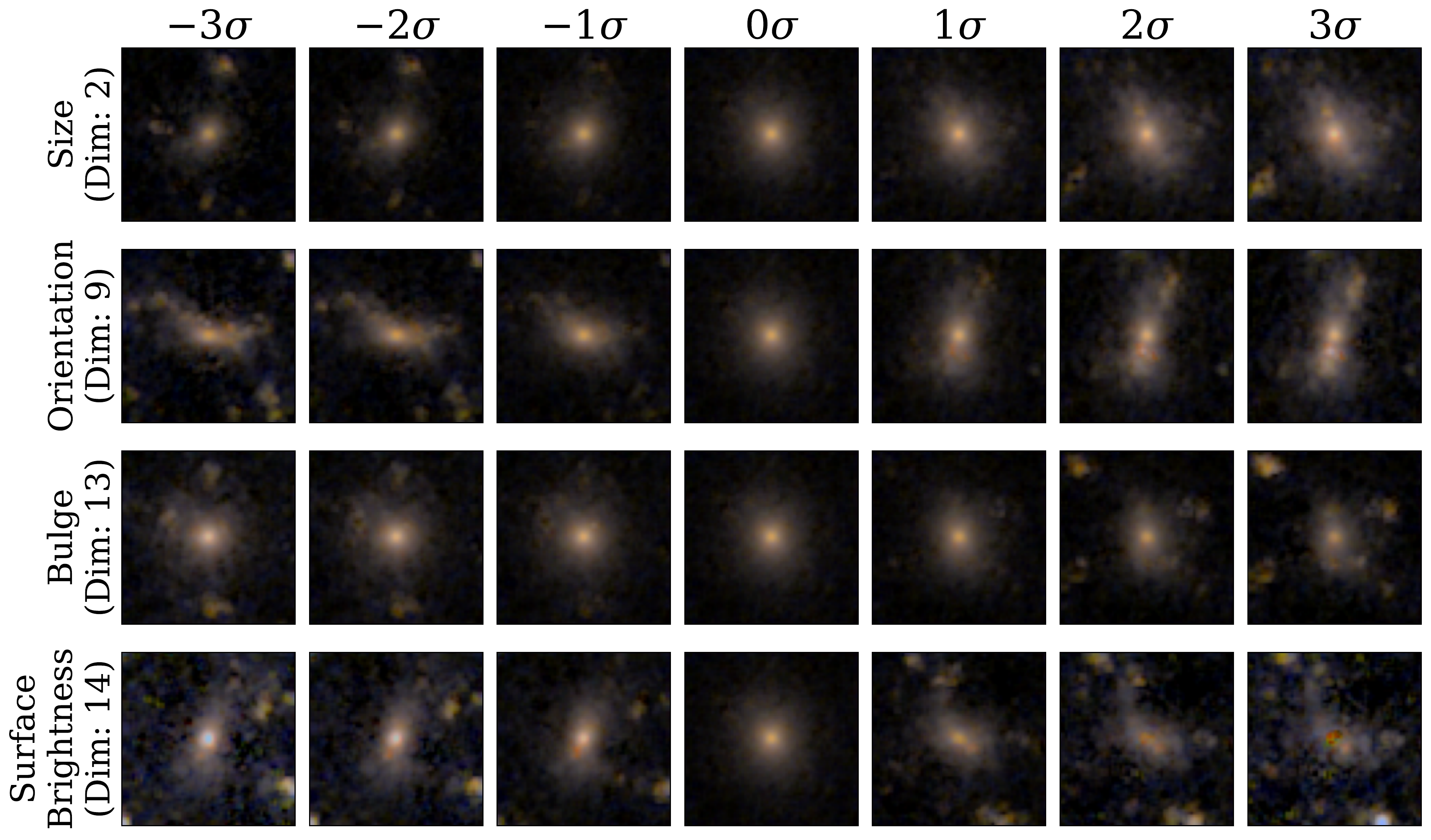}
  \caption{The first elliptical galaxy from Fig.~\ref{fig:recon}}
  \label{fig:tinker_spheroid}
  \end{subfigure}
  \caption{Synthetic galaxy images generated by perturbing capsule dimensions. Each column shows the decoded image when one of the 16 dimensions of the capsule vector is perturbed in units of its standard deviation (keeping all the others fixed). The $0\sigma$ column shows the decoded image from the unperturbed capsule and are identical for each row. We show a subset of the dimensions here for which the perturbations have a clear interpretation (see Appendix~\ref{appendix:tinker} for a version with all the dimensions). We see that some of the capsule dimensions, encode physical features like size, orientation, amount of central bulge and surface brightness of the galaxies.}
\label{fig:tinker}
\end{figure*}

Fig.~\ref{fig:tinker} shows the synthetic galaxy images generated from the perturbed capsule vectors for two galaxies (the first instance of each morphological type from Fig.~\ref{fig:recon}). We can see that perturbing specific components change properties like size (i.e., the angular size of the galaxy and how fast the light profile falls off), orientation, amount of central bulge, and surface brightness. This shows that some of the features learnt by the capsule network correspond to physical properties of galaxies. Visual properties like size and surface brightness change with the distance of the galaxies and can help to break degeneracies in the colour-redshift relation and provide better redshift inference. Fig.~\ref{fig:tinker} shows the synthetic images from perturbed capsules for only a subset of dimensions for which the change in the images is easily identifiable visually. Appendix \ref{appendix:tinker} shows the synthetic images generated by perturbing all 16 of the dimensions individually.

\subsubsection{Correlations of Capsule Dimensions with Physical Properties}\label{sec:dist-corr}
\footnotetext{\url{https://biprateep.github.io/encapZulate-1/viz/explore_UMAP_DenseMAP.html}}

To check whether any physical properties of the galaxies are encoded by the capsules that cannot be identified by simply looking at synthetic images generated from perturbed capsules, we measure the correlations between each dimension of the capsules and various global galaxy properties. Since we expect the correlations to be non-linear in nature, we use the distance correlation \citep{Szekely2007DistCorr} to measure them. The distance correlation quantifies the dependence between two random variables by measuring how much the Euclidean distance between two samples of one random variable changes for a given change in distance between two samples of another random variable. This makes the distance correlation sensitive to any kind of dependence between two random variables, unlike Pearson or Spearman correlations which measure linear and strictly monotonic relationships respectively. The distance correlation has a value between $0$ and $1$, where $0$ would mean that the random variables are independent whereas a value of $1$ would mean the linear sub-spaces spanned by the two random variables are almost equal, indicating a very high degree of dependence.

\begin{figure*}
 \includegraphics[width=\linewidth]{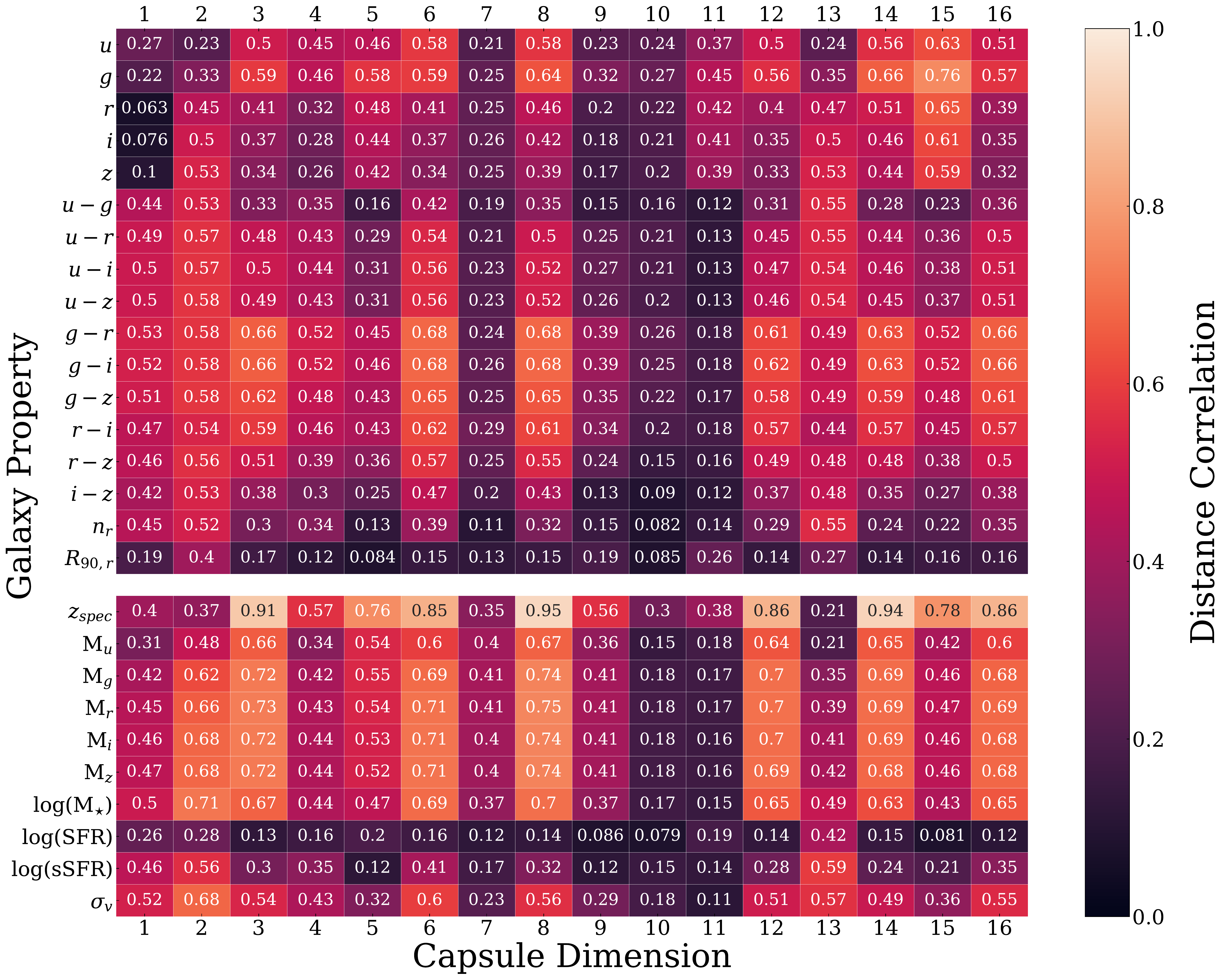}
 \caption{Measurements of the distance correlation between the capsule dimensions corresponding to the predicted morphological class and global galaxy properties (as described in Sec.~\ref{sec:data-galaxy-prop}). The values have a range between 0 and 1 where a value of 0 means the two random variables being compared are independent and a value of 1 indicates a high level of dependency. We have grouped the galaxy properties into two sets: properties which solely depend on photometry (top) and properties which include knowledge of the spectroscopic redshift along with photometry (bottom). $u,g,r,i,z$ represent the extinction corrected cmodel magnitudes. $u-g$, $g-r$, etc. represent galaxy colours calculated using extinction corrected model magnitudes. $n_{r}$ and $R_{90,r}$ represent the S\'ersic index and the 90\% light radius obtained from the S\'ersic profile fit to $r$-band photometry and are used as a proxy for a galaxy's size. $z_{\mathrm{spec}}$ denotes the spectroscopic redshift; $\mathrm{M}_{u/g/r/i/z}$ represent the absolute magnitudes in each of the five bands.  $\mathrm{M}_{\star}$ stands for the stellar mass, SFR stands for the star formation rate, and sSFR stands for the specific star formation rate. $\sigma_{v}$ represents the velocity dispersion of the spectra. We see that the components of the capsule vectors are not only correlated with the spectroscopic redshift but also correlated with the apparent magnitudes and colours, measurements that are traditionally used by photometric redshift prediction algorithms. We also see that they are well correlated with parameters of a S\'ersic fit which are indirect indicators of morphology as well as physical properties of the galaxies that would traditionally require spectroscopic measurements.}
 \label{fig:correlations}
\end{figure*}

Fig.~\ref{fig:correlations} shows values of distance correlation between each of the components of the capsule vector corresponding to the predicted morphology and global properties of galaxies in the test set. Unsurprisingly, we find that many of the capsule components have strong correlations with the spectroscopic redshift, with dimensions 8, 14 and 3 being the strongest. The capsule dimensions that show strong correlations with spectroscopic redshift also show strong correlations with observed frame galaxy colours and apparent magnitudes which are known to be good predictors of photometric redshift. Given this pattern, we also expect them to be well correlated with galaxy absolute magnitudes ($\mathrm{M}_{u/g/r/i/z}$) which we can also verify from Fig.~\ref{fig:correlations}. S\'ersic index ($\mathrm{n}_{r}$) correlates the most with dimension 13 which we saw controls the amount of a galaxy's central bulge (see Fig.~\ref{fig:tinker}). Similarly, dimension 2 which we saw control the visual size of the galaxy image has the strongest correlation with the 90\% light radius ($R_{90,r}$) among all capsules and also correlates well with S\'ersic index which are the two quantities which together quantify the visual size of the galaxy on the sky. We can therefore infer that the capsules successfully encode almost all of the photometric properties of the galaxy image. A few illustrative examples of these correlations in form of scatter plots can be found in appendix~\ref{appendix:scatter}.

Many capsule dimensions show correlation with physical properties like stellar mass ($\mathrm{M}_{\star}$) and velocity dispersion of the spectra ($\sigma_{v}$) and a small number of dimensions show strong correlations with star formation rate ($\mathrm{SFR}$) and specific star formation rate ($\mathrm{sSFR}$). Most likely, these correlations arise because SFR and sSFR depend on galaxy magnitudes and spectroscopic redshifts which the capsules efficiently encode, but the capsules may also encode some physical properties of the galaxies. Even though we focus on predicting photometric redshifts in this work, we expect that capsule-based encodings can be used to create a general purpose image-based inference methodology for physical properties of galaxies and will be explored in a future work.

\subsubsection{Feature Importance Using SHAP Values}
As shown in the Secs.~\ref{sec:tinker}~and~\ref{sec:dist-corr}, each capsule dimension tends to encode a somewhat different property of the input image, so we would like to see which of the dimensions are most useful in predicting photo-$z$'s. To quantify this, we calculate the SHapley Additive exPlanations (SHAP; \citealt{Lundberg2017SHAP}) values for each of the capsule dimensions that are used by the redshift prediction network using the test data. SHAP is a method to explain a prediction by computing the contribution of each feature. It takes a game theory approach to optimally distribute credit to each feature for a given prediction using Shapley Values \citep{Shapley1953ShapleyValue}. The Shapley value for a feature is defined as the average marginal contribution of a feature across its all possibilities for a given prediction. The SHAP value is then calculated via a weighted sum of Shapley values to ensure that the contribution of each feature to a prediction add up to the value of the prediction. Since it would be prohibitively expensive to calculate contributions across all possibilities of the feature space, we use the expected gradients method which combines ideas from Integrated Gradients \citep{Sundararajan2017IntegratedGradients}, SHAP \citep{Lundberg2017SHAP} and SmoothGrad \citep{Smilkov2017SmoothGrad} to approximately calculate the SHAP values for a neural network. A positive SHAP value indicates that the particular value of the feature increases the value of the output, a negative SHAP value indicates that the output is decreased, whereas a value of zero means that the feature does not contribute towards the output for that specific prediction. We then rank the features (i.e., capsule dimensions) based on their magnitude of SHAP values averaged over all predictions in the test set. Thus, a capsule dimension is deemed to be the most important if it influences the output most across all the predictions.  

\begin{figure}
  \includegraphics[width=1\linewidth]{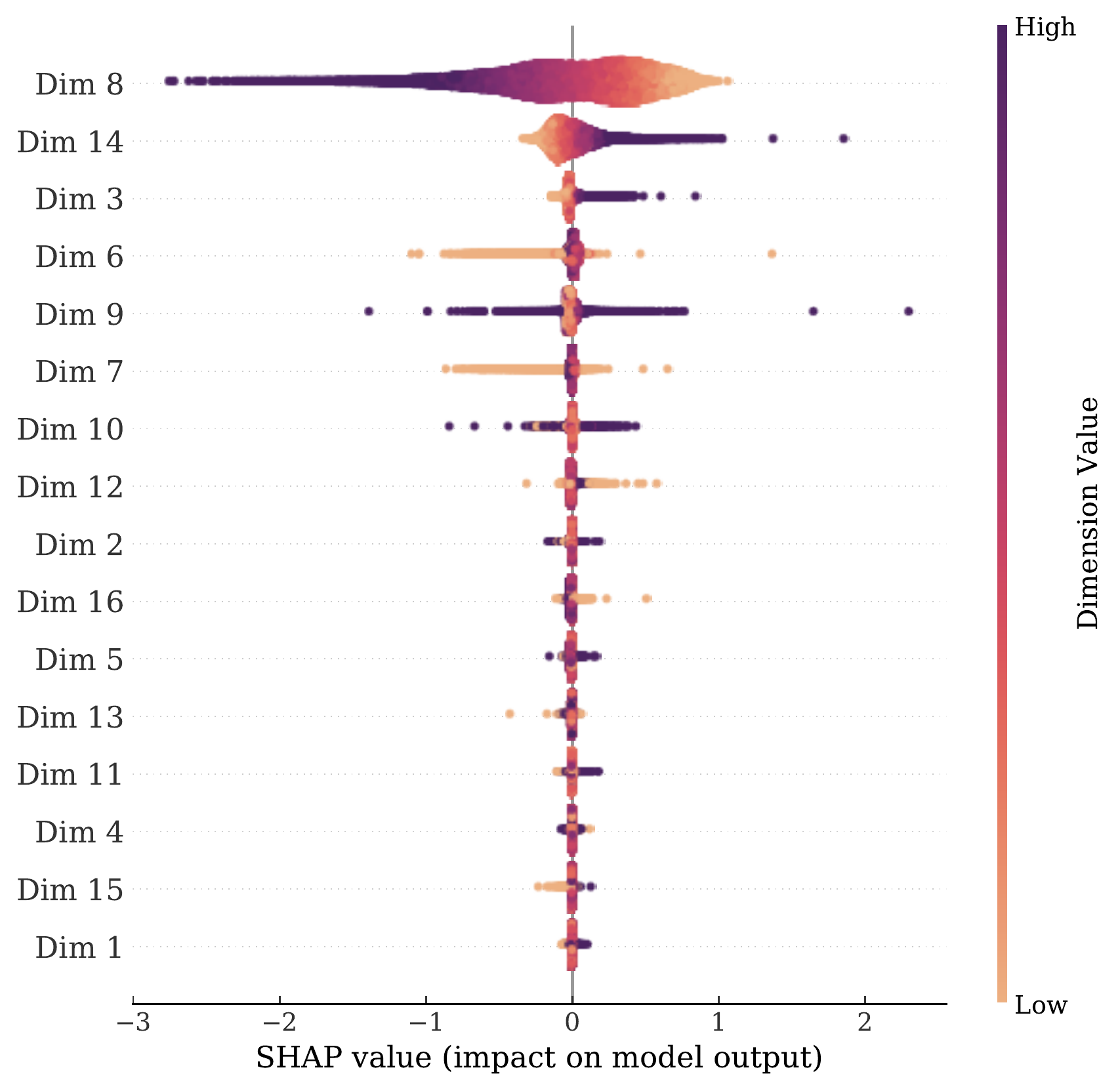}
  \caption{A SHAP summary plot showing the SHAP values of each capsule dimension for the entire test set. The capsule dimensions are listed in decreasing order of their importance (i.e. average magnitude of SHAP values). The points are colour-coded as per the value of the capsule dimension. We see that dimensions 8 and 14 are the most important followed by dimensions 3 and 6. The pile-up of points at a SHAP value of zero indicates that the dimension does not contribute towards the prediction for this specific case and the network gets similar information from another capsule dimension. This can happen when features are correlated. We do see that all the dimensions have some non-zero SHAP values, indicating that all the dimensions contribute towards the prediction at least sometimes.}
  \label{fig:feature-importance}
\end{figure}

We show the SHAP values for each prediction in the test set in the summary plot shown in Fig.~\ref{fig:feature-importance}. The capsule dimensions are listed in decreasing order of their importance (i.e., average magnitude of SHAP values). The points are also colour coded as per the value of the feature which helps us to qualitatively identify how much the prediction changes based on a change in the value of the dimension. We see that capsule dimensions 8 and 14 are the most important, followed by dimensions 3 and 6. The next four capsule dimensions still contribute significantly to the prediction as dimensions 9, 7, 10, 12 have relatively high SHAP values. All the other dimensions contribute to the prediction significantly only a small number of times. For many of the dimensions, we see a pile up of SHAP values around 0. This indicates that the particular feature does not contribute much towards the prediction for that specific case. This can happen if the input features are correlated and the model gets similar information from a different dimension for that specific prediction. This is also evident from the fact that the 2D UMAP projection of the capsules form a nearly perfect redshift sequence (see Fig.~\ref{fig:umap}) suggesting that the data do not fully span the 16-dimensional latent space. We therefore define the importance ranking of a capsule dimension  as an average over the entire test set and the ranking may be different for a specific prediction.

Dimension 8 has the highest SHAP feature importance. Although we cannot clearly discern what physical property it represents from the synthetic images generated from perturbed capsules, we can see from the figures in appendix~\ref{appendix:tinker} that perturbing this dimension causes the image to morph from an elliptical galaxy to a spiral galaxy. We hypothesise that dimension 8 learns a representation which is a combination of the morphological type, colour, and orientation of the galaxy which helps it to distinguish between an elliptical galaxy which is intrinsically red and an edge-on spiral galaxy which appears to be reddened because of dust. This helps the capsule network to learn representations of galaxy colour while being aware of the morphology and orientation which can be very useful to break degeneracies in the colour-redshift relation. We also see that dimension 14 is the second most important feature. Fig.~\ref{fig:tinker} shows that dimension 14 encodes information about the surface brightness of the observed galaxy. A lower value of dimension 14 corresponds to a brighter object. From Fig.~\ref{fig:feature-importance} we see that a lower value of dimension 14 reduces the redshift prediction since they have a negative SHAP value. This shows that the neural network assigns a lower redshift to objects with higher surface brightness. Surface brightness is a very good proxy to the distance of a galaxy (and therefore redshift) since objects farther away appear fainter at a fixed luminosity. Learning a representation of surface brightness hence helps the network to better predict redshifts.

\section{Summary and Discussion} \label{sec:summary}
In this paper, we use a deep capsule network to produce photometric redshift point estimates from images of galaxies and provide interpretation of the features learnt by the network. We use $\sim 400,000$ SDSS $ugriz$ images, their spectroscopic redshifts, and morphological class labels from Galaxy-Zoo-1 (see Sec.~\ref{sec:data}) to train our deep capsule network. Capsule networks are a new type of neural network architecture that are better suited for identifying morphological features than traditional CNNs. We use a deep capsule network architecture that uses 3D convolution based routing mechanisms and skip connections to efficiently train the network (see Sec.~\ref{sec:capsnets} and Fig.~\ref{fig:network-arch}).

We achieve a photometric redshift prediction accuracy comparable to or better than current methods while requiring less data and fewer trainable parameters (see Figs.~\ref{fig:point-est} \& \ref{fig:performance_v_data}). The performance of our algorithm is stable across the brightness and redshift range of our data set (see Fig.~\ref{fig:met_v_z}). Moreover, the decision-making of our capsule network is easier to interpret as capsules act as a low-dimensional encoding of the input image and can be used to produce reconstructed images (see Fig.~\ref{fig:recon}). We use UMAP, a non-linear dimensionality reduction method to embed the capsules in 2-dimensional space and show that the capsules produce an almost perfect redshift sequence with the fraction of spirals in a region exhibiting a gradient roughly perpendicular to the redshift sequence (see Fig.~\ref{fig:umap}). We then perturb the encodings of real galaxy images to generate synthetic galaxy images that demonstrate the image properties (e.g., size, orientation, and surface brightness) encoded by each capsule dimension (see Fig.~\ref{fig:tinker}). We calculate the feature importance of each capsule dimension using their SHAP values to rank them based on their usefulness towards predicting photo-$z$'s (see Fig.~\ref{fig:feature-importance}). We also demonstrate that galaxy properties (e.g., magnitudes, colours, and stellar mass) correlate strongly with each capsule dimension (see Fig.~\ref{fig:correlations}). This tells us that the capsule dimensions encode and use visual and morphological properties of galaxy images (like surface brightness, orientation) in addition to measures of amount of light (like colours and magnitudes) to infer the photometric redshift.

Here we have presented photo-$z$ point estimates, though for many science cases photo-$z$ PDFs are more desirable and sometimes necessary for meaningful analyses. However, current ML-based photo-$z$ PDF estimation efforts suffer from poor calibration \citep{Schmidt2020LSSTPhotoz}. In future work, we plan to incorporate methods described in \citet{Dey2021Recalibration,Dey2022Recal} to properly calibrate ML-based photo-z PDFs based on a galaxy's position in input space with  capsule network photo-$z$ PDFs serving as a natural example to demonstrate the expected improvements.

More generally, the future of capsule network-based photo-$z$ estimation looks bright. Their high training efficiency will allow for deeper and wider models with greater capacity to handle the massive training sets from current and future spectroscopic surveys like DESI \citep{DESIEtal2016SurveyDescription} and PFS \citep{Takada2014PFSplan} that extend to higher redshifts, span a wider redshift range, and probe to fainter magnitudes. Specifically, we plan to enable early DESI science by estimating photo-$z$'s for objects in the DESI Legacy Imaging Surveys \citep{DeyEtal2019DesiLegacySurvey} before the DESI spectroscopic survey is complete. At even higher redshifts, we are optimistic that capsule networks can leverage morphology---especially the evolution of galaxy morphologies from $z \sim 2$ to $z < 0.5$---from space-based high-resolution imaging to help break the SED degeneracies that plague template-fitting methods at high-$z$.  With growing high-$z$ spectroscopic training sets and rapidly progressing capsule network architecture development, we are optimistic that capsule networks will provide complementary constraints or even superior photo-$z$'s to template-based methods at high-$z$.

\section*{Acknowledgements}
BD, BHA and JAN acknowledge the support of the National Science Foundation under Grant No. AST-2009251. Any opinions, findings, and conclusions or recommendations expressed in this material are those of the author(s) and do not necessarily reflect the views of the National Science Foundation. Support for YYM was provided by NASA through the NASA Hubble Fellowship grant no.\ HST-HF2-51441.001 awarded by the Space Telescope Science Institute, which is operated by the Association of Universities for Research in Astronomy, Incorporated, under NASA contract NAS5-26555. RZ is supported by the Director, Office of Science, Office of High Energy Physics of the U.S. Department of Energy under Contract No. DE-AC02-05CH11231. Argonne National Laboratory's work was supported by the U.S. Department of Energy, Office of Science, under contract DE-AC02-06CH11357.

The pre-processed galaxy images and their spectrscopic redshifts used in this work were compiled by Emmanuel Bertin and sent to us by Johanna Pasquet and Marie Treyer. We are indebted to them for providing us early access to the data set and providing relevant documentation.

This research was supported in part by the University of Pittsburgh Center for Research Computing through the resources provided. We specifically acknowledge the assistance of Barry Moore II.

The authors would like to thank Rachel Bezanson, Sonja Cwik, Scott Dodelson, Ayres Freitas, Yasha Kaushal, Ann Lee, Christine Mazzola Daher, Alan Pearl and David Setton for helpful discussions and suggestions during the course of this work. The authors would also like to thank the anonymous reviewer for their helpful comments and suggestions.

This research made use of the following software packages: Astropy \citep{AstropyEtal2013, AstropyEtal2018}, Colorcet \citep{Kovesi2015Colorcet}, Keras \citep{Chollet2015Keras},  Matplotlib \citep{Hunter2007Matplotlib}, Numpy \citep{CharlesEtal2020Numpy}, Pandas \citep{mckinney2010Pandas,reback2020Pandas}, Scikit-Learn \citep{PedregosaEtal2011Sklearn}, Scipy \citep{VirtanenEtal2020Scipy}, Seaborn \citep{Waskom2021Seaborn}, SHAP \citep{Lundberg2017SHAP}, Tensorflow \citep{AbadiEtal2015Tensorflow}, UMAP \citep{Mcinnes2018UMAPSoftware} and WebPlotDigitizer \citep{Rohatgi2020WebplotDigitizer}.

Funding for SDSS-III has been provided by the Alfred P. Sloan Foundation, the Participating Institutions, the National Science Foundation, and the U.S. Department of Energy Office of Science. The SDSS-III website is http://www.sdss3.org/.

SDSS-III is managed by the Astrophysical Research Consortium for the Participating Institutions of the SDSS-III Collaboration including the University of Arizona, the Brazilian Participation Group, Brookhaven National Laboratory, Carnegie Mellon University, University of Florida, the French Participation Group, the German Participation Group, Harvard University, the Instituto de Astrofisica de Canarias, the Michigan State/Notre Dame/JINA Participation Group, Johns Hopkins University, Lawrence Berkeley National Laboratory, Max Planck Institute for Astrophysics, Max Planck Institute for Extraterrestrial Physics, New Mexico State University, New York University, Ohio State University, Pennsylvania State University, University of Portsmouth, Princeton University, the Spanish Participation Group, University of Tokyo, University of Utah, Vanderbilt University, University of Virginia, University of Washington, and Yale University.

\section*{Data Availability}
The images, spectroscopic redshifts and associated value added catalogues from \citet{PasquestEtal2019Photoz} used in this work are available at \href{https://deepdip.iap.fr/\#item/60ef1e05be2b8ebb048d951d}{https://deepdip.iap.fr/\#item/60ef1e05be2b8ebb048d951d}. Python code used to train and test the models, additional value added catalogues and redshift predictions by our model along with interactive visualizations are available at \href{https://biprateep.github.io/encapZulate-1/}{https://biprateep.github.io/encapZulate-1/}.


\bibliographystyle{mnras}
\bibliography{astro_references,ml_references,software_references} 


\newpage
\appendix

\section{Capsule Networks and Routing Mechanisms}\label{appendix:routing}
To construct the classification-and-encoding network, we first use a set of convolutional filters, the outputs of which are reshaped into a set of tensors which are treated as the initial set of capsules. We then use two main kind of capsule layers, dynamic routing-based class capsules and convolution routing-based capsules (i.e., \texttt{Conv-Caps} and \texttt{3D-Conv-Caps} layers in Fig.~\ref{fig:network-arch}). As shown in Fig.~\ref{fig:network-arch}, the convolution routing based capsules are used to construct the hidden layers whereas the dynamic routing based capsules are used to construct the output layer of the classification-and-encoding network where each capsule represents a morphological type. In this section we give a brief overview of the mathematical aspects of the capsule layer architectures used in this work. This is intended to be a short summary and interested readers are recommended to refer to \citet{SabourEtal2017Capsnet} for a detailed discussion on capsules with dynamic routing and \citet{RajasegaranEtal2019Deepcaps} for convolutional capsules. We have tried to follow the same mathematical notation used by these two works for easy reference. 

\subsection{Dynamic Routing (i.e. Routing by Agreement)}\label{appendix:dynamicrouting}
Let $\mathbf{u}_{i}$ denote the $i^{\mathrm{th}}$ capsule vector in layer $l$ of the network and $\mathbf{v}_{j}$ denote the $j^{\mathrm{th}}$ capsule vectors in layer $l+1$. To obtain the capsules in layer $l+1$ from the ones in layer $l$ we define an intermediate ``prediction'' vector ($\mathbf{\hat{u}}_{j|i}$) as:
\begin{equation}
\mathbf{\hat{u}}_{j|i}=\mathbf{W}_{ij}\mathbf{u}_{i},
\end{equation}
where $\mathbf{W}_{ij}$ is a weight matrix learnt by gradient descent. The capsules in the following layer ($\mathbf{v}_{j}$) are calculated using a weighted sum of these prediction vectors after being passed through a non-linear activation function called the squashing function defined as:
\begin{equation}
    \mathbf{v}_j = \frac{\lVert \mathbf{s}_j \rVert^2}{1 + \lVert \mathbf{s}_j \rVert^2}\frac{\mathbf{s}_j}{\lVert \mathbf{s}_j \rVert},
\end{equation}
where $\mathbf{s}_j$ is the weighted sum given by:
\begin{equation}
    \mathbf{s}_j = \sum_{i}c_{ij}\mathbf{\hat{u}}_{j|i},
\end{equation}
where $c_{ij}$ are the coupling coefficients determined by an iterative process. To ensure that they always add up to 1, they are defined in terms of the softmax transformed variables $b_{ij}$ as:
\begin{equation}
    c_{ij} = \frac{\exp(b_{ij})}{\sum_{k}\exp(b_{ik})}.
\end{equation}
The variables $b_{ij}$ can be treated as the log prior probability that the capsule $i$ in layer $l$ is coupled to the capsule $j$ in layer $l+1$. In a single pass of back propagation, we begin with $b_{ij}=0$ to provide equal weights to all the capsules initially, and then the coupling coefficients are iteratively updated by measuring the agreement between the current output of each capsule in layer $l+1$, i.e., $\mathbf{v}_{j}$ and the prediction made by the capsules in layer $l$, i.e., $\mathbf{\hat{u}}_{j|i}$. The agreement is defined as the scalar product $\mathbf{v}_{j}.\mathbf{\hat{u}}_{j|i}$ and is added to $b_{ij}$ before computing the coupling coefficients. So, for each step in the iteration:
\begin{equation}
    b_{ij}\leftarrow b_{ij}+\mathbf{v}_{j}.\mathbf{\hat{u}}_{j|i}.
\end{equation}
The number of iterations is a tunable hyper-parameter. Larger number of iterations will provide better estimates of the coupling coefficients at the cost of increasing the number of computations. We use 3 iterations as it was found to work reasonably well by \citet{SabourEtal2017Capsnet} who proposed this algorithm. Since these capsules ($\mathbf{v}_{j}$) form the final layer of the classification-and-encoding network, we calculate their Euclidean norms which are used as a measure of the class probabilities the capsules represent. These predicted class probabilities are then used as inputs to the margin loss function (eq.~\ref{eq:margin-loss}).

\subsection{Convolution based routing}
One of the drawbacks of the dynamic routing algorithm described in Sec.~\ref{appendix:dynamicrouting} is that the computations are done in a way analogous to fully connected neural networks. This means that the number of trainable weights increase dramatically for a deep network architecture required for complex tasks like predicting photo-$z$'s. To solve this problem, \citet{RajasegaranEtal2019Deepcaps} proposed capsule network layers that use computationally efficient convolutional operations. We use them as the intermediate layers of our classification-and-encoding network. The weights of the convolution filters are determined using gradient descent whereas the coupling coefficients for routing are determined by an iterative process. In the initial layers, the feature maps obtained from convolution operations is large and iterative routing can be expensive. So, following \citet{RajasegaranEtal2019Deepcaps}, we use a mix of two kinds of convolutional capsule layers, one which does one routing iteration (viz. \texttt{Conv-Caps}) and another one doing 3 routing iterations (viz. \texttt{3D-Conv-Caps}) in our network architecture (see Fig.~\ref{fig:network-arch}).

To facilitate convolution operations, the capsules start out as 3D tensors which are flattened into 1D capsule vectors when we reach the final layer in our architecture. Let the output of the convolutional capsule layer $l$ be $\mathbf{\Phi}^{l}\in \mathbb{R}^{(w^{l}, w^{l}, c^{l}, n^{l})}$, where $w^{l}$ denotes the height and width, $c^{l}$ the depth, and $n^{l}$ the number of 3D capsule tensors. Similarly, let $\mathbf{\Phi}^{l+1}\in \mathbb{R}^{(w^{l+1}, w^{l+1}, c^{l+1}, n^{l+1})}$ represent the output of the layer $l+1$. 

The \texttt{Conv-Caps} layer first reshapes $\mathbf{\Phi}^{l}$ into a tensor of shape $(w^{l}, w^{l}, c^{l}\times n^{l})$ and convolves with $(c^{l+1}\times n^{l+1})$ number of filters, producing $(c^{l+1}\times n^{l+1})$ number of feature maps of shape $(w^{l+1}, w^{l+1})$. They are then reshaped into a tensor of shape $(w^{l+1}, w^{l+1}, c^{l+1}, n^{l+1})$. This 3D tensor ($S_{pqr}$) is then used as the input to a nonlinear squashing function defined by:
\begin{equation}
    \hat{S}_{pqr} = \frac{\lVert S_{pqr} \rVert^2}{1 + \lVert S_{pqr} \rVert^2}\frac{S_{pqr}}{\lVert S_{pqr} \rVert} \label{eq:squash3d}.
\end{equation}
Since we will use just one iteration of routing for this layer, the output of the squashing function is treated as the output of the layer (i.e., $\mathbf{\Phi}^{l+1}= \mathbf{\hat{S}}$).

For the \texttt{3D-Conv-Caps} layer, we first reshape $\mathbf{\Phi^{l}}$ into a tensor of shape $(w^{l}, w^{l}, c^{l}\times n^{l}, 1)$. Then, it is convolved with $(c^{l+1}\times n^{l+1})$ number of 3D convolution kernels of appropriate shape so as to produce a tensor of shape $(w^{l+1},w^{l+1},c^{l}, c^{l+1}\times n^{l+1})$. It is then reshaped into a tensor, $\mathbf{\Tilde{V}}$ of shape $(w^{l+1},w^{l+1},c^{l},n^{l+1}, c^{l+1})$ which acts as the intermediate ``prediction'' tensor. The capsules of the following layer are then calculated via the weighted sum of tensors given by:
\begin{equation}
    S_{pqr} = \sum_{s}k_{pqrs}.\Tilde{V}_{pqrs}.
\end{equation}
Then $\mathbf{S}$ is used as an input to the tensor squashing function defined in eq.~\ref{eq:squash3d} to obtain the squashed tensor, $\mathbf{\hat{S}}$ which after the iterative updates will be treated as the output capsules($\mathbf{\hat{\Phi}}^{l+1}$). The coupling coefficients for the weighted sum ($k_{pqrs}$) are determined by an iterative process. To ensure that they are normalised they are defined in terms of softmax transformed variable $\mathbf{B}\in \mathbb{R}^{(w^{l+1},w^{l+1},c^{l+1},c^{l})}$ given by:
\begin{equation}
    k_{pqrs}=\dfrac{\exp{(b_{pqrs})}}{\sum_{x}\sum_{y}\sum_{z}\exp{(b_{xyzs})}}.
\end{equation}

In a single pass of back propagation we begin with $b_{pqrs}=0$ to provide equal weight to all capsules initially and then the coupling coefficients are iteratively updated 3 times by measuring the agreement (via the scalar product) between the current output of the capsules and the intermediate prediction tensors in each iteration i.e.,
\begin{equation}
    b_{pqrs}\leftarrow b_{pqrs} + \hat{S}_{pqr}.\Tilde{V}_{pqrs}.
\end{equation}

Finally, when the output of the convolutional capsules are used as inputs to the capsules with dynamic routing, the tensors in a layer $l$ of shape $(w^{l}, w^{l}, c^{l}, n^{l})$ are flattened to the shape $(w^{l} \times w^{l} \times c^{l}, n^{l})$, i.e. we get $n^{l}$ number of capsule vectors each with $w^{l} \times w^{l} \times c^{l}$ number of dimensions.

\section{Synthetic Images from perturbed capsule components}\label{appendix:tinker}
Here we show an extended version of Fig.~\ref{fig:tinker} with synthetic galaxy images generated from perturbing all 16 of the dimensions individually. Each column shows the decoded image when one of the 16 dimensions of the capsule vector is perturbed in units of its standard deviation (keeping all the others fixed). The $0\sigma$ column shows the decoded image from the unperturbed capsule and are identical for each row. Since the capsule network training process does not disentangle the features learnt by each dimension, not all the dimensions control a single easily identifiable feature. A subset of the dimensions for which the the features are easily identifiable are shown in Fig.~\ref{fig:tinker}.

\begin{figure*}
  \begin{subfigure}[b]{0.49\textwidth}
    \includegraphics[width=\textwidth]{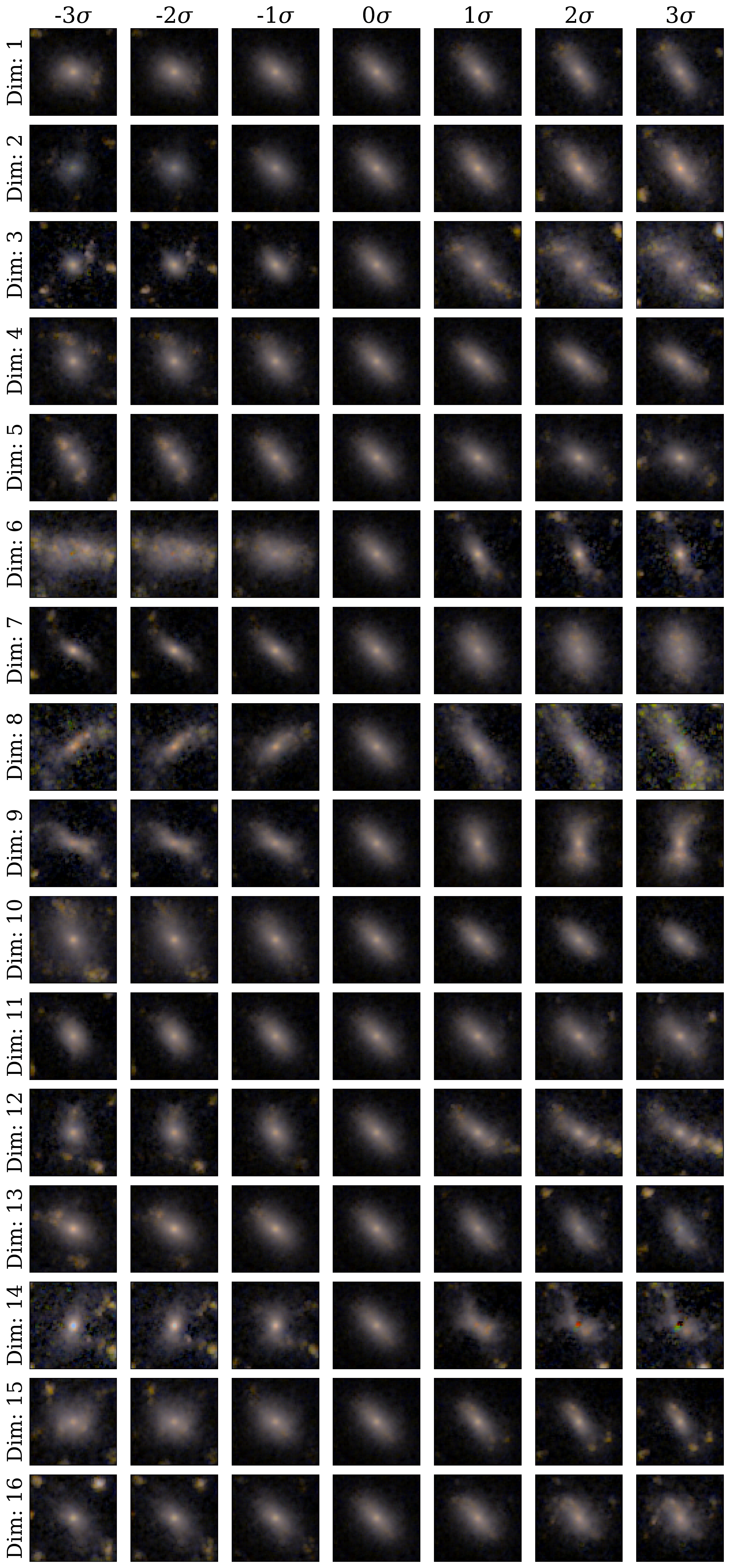}
    \caption{The first spiral galaxy from Fig.~\ref{fig:recon}}
    \label{fig:disks-appendix}
  \end{subfigure}
  \begin{subfigure}[b]{0.49\textwidth}
    \includegraphics[width=\textwidth]{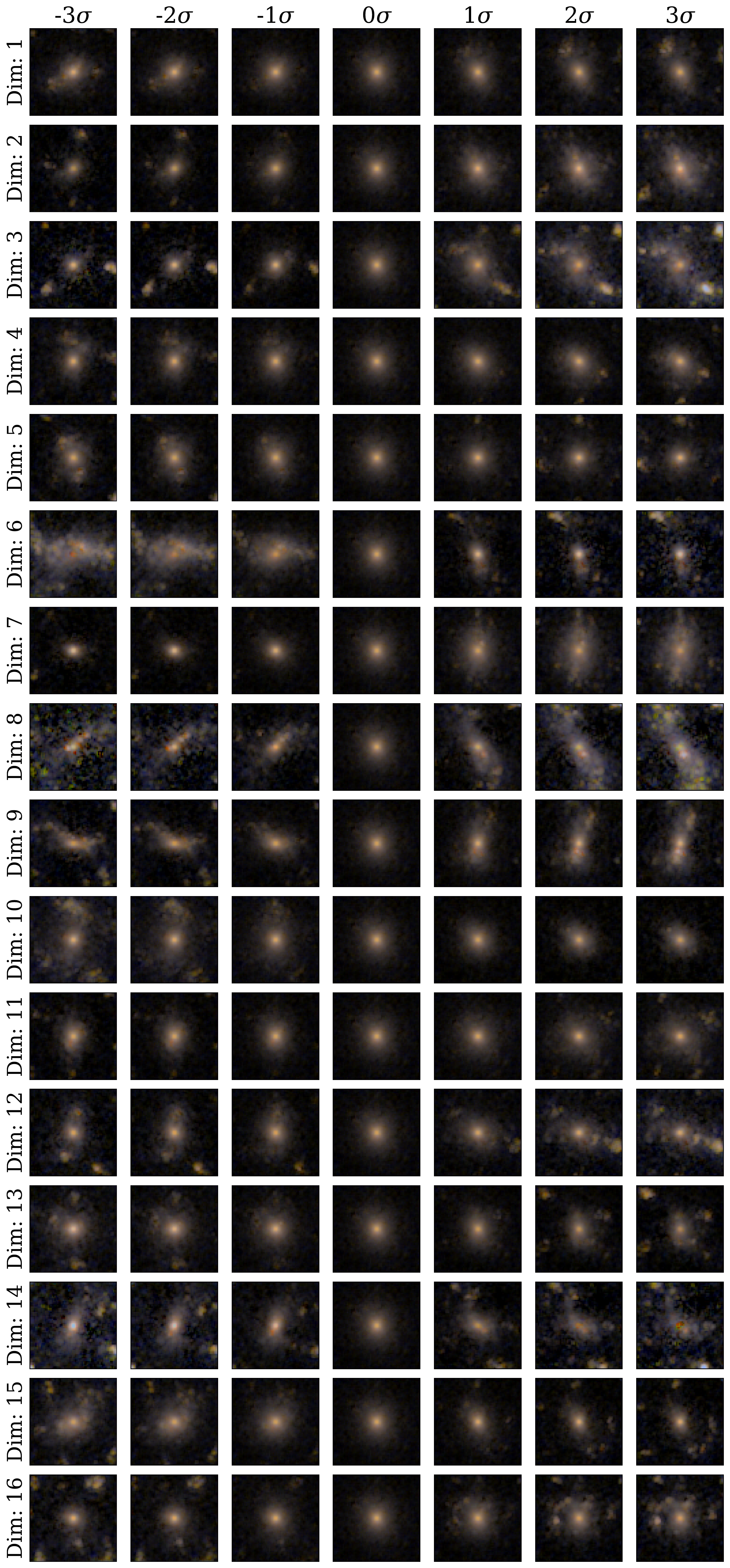}
    \caption{The first elliptical galaxy from Fig.~\ref{fig:recon}}
    \label{fig:spheroids-appendix}
  \end{subfigure}
\caption{Reconstructions from perturbed capsule vectors. Each column shows the reconstructions when one of the 16 components of the capsule vector is perturbed in units of their standard deviation (keeping all the others fixed). This is an extended version of the Fig.~\ref{fig:tinker} and shows reconstructions from perturbations of all the dimensions.}
\end{figure*}

\section{Correlations of Capsule Dimensions with Physical Properties} \label{appendix:scatter}
\begin{figure*}
\includegraphics[width=\textwidth]{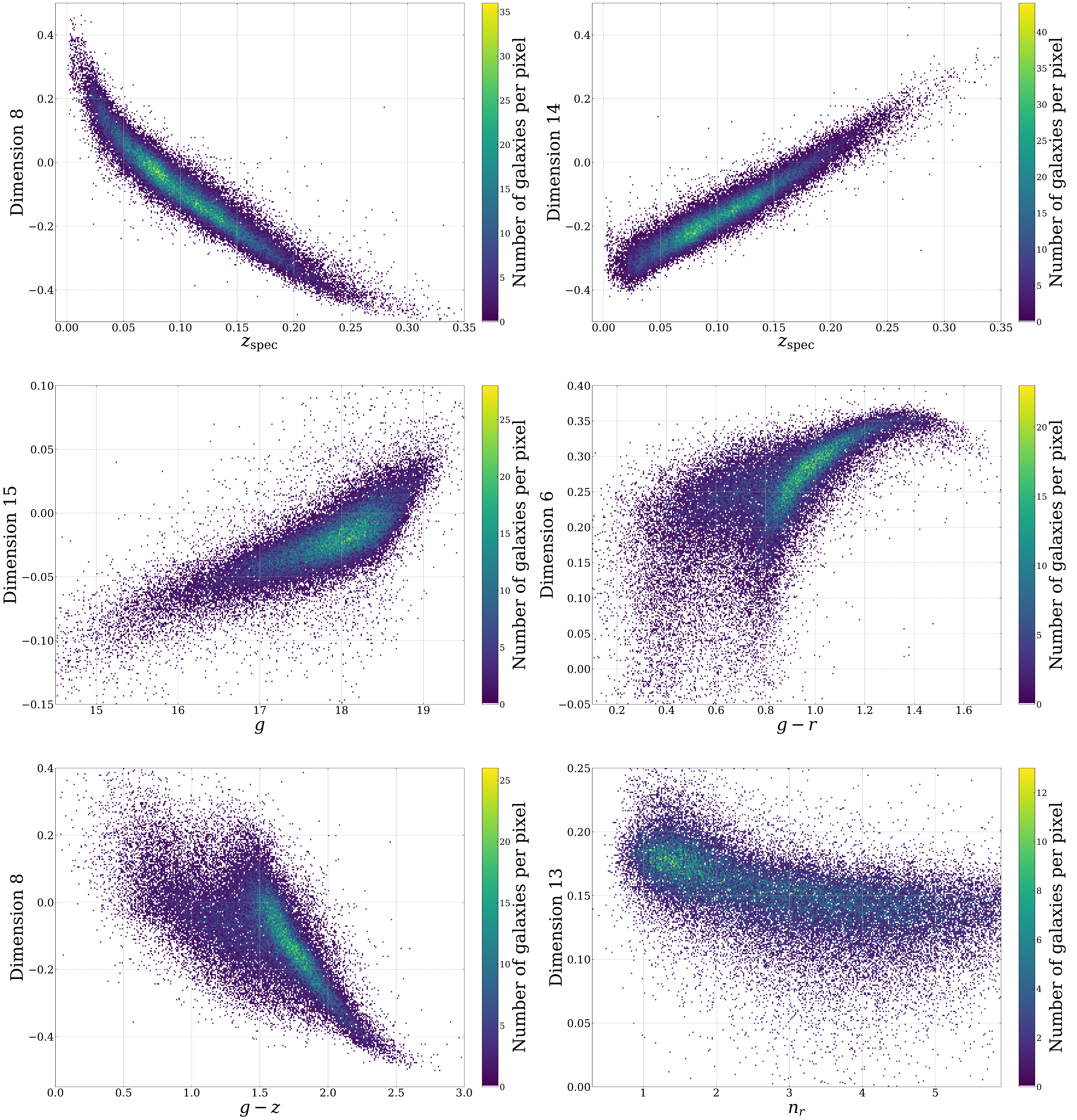}  
\caption{A few examples of strong correlations between capsule dimensions and physical properties of galaxies visualised using scatter plots. $g$ represents the extinction corrected SDSS $g$-band cmodel magnitude. $g-r$ and $g-z$ represent galaxy colours calculated using extinction corrected model magnitudes. $n_{r}$ represents the S\'ersic index obtained from a S\`ersic profile to the $r$-band photometry. We observe that the value of the capsule dimensions varies with the galaxy property indicating some correlation.}
\label{fig:caps-scatter}
\end{figure*}

A few illustrative examples of strong correlations between capsule dimensions and physical properties of galaxies have been visualised using scatter plots in Fig.~\ref{fig:caps-scatter}. We observe that the value of the capsule dimensions varies with the galaxy property indicating some correlation.

\bsp	
\label{lastpage}
\end{document}